\documentclass[twocolumn]{aastex631}

\usepackage{multirow}
\usepackage{soul}
\usepackage{xspace}

\newcommand{\bpic}{$\beta$~Pic\xspace}
\newcommand{\Bpic}{$\beta$~Pic\xspace}
\newcommand{\BPic}{$\beta$~Pic\xspace}
\newcommand{\um}{\text{\textmu m}\xspace}

\begin{document}

\title{JWST-TST High Contrast: Asymmetries, dust populations and hints of a collision in the $\beta$ Pictoris disk with NIRCam and MIRI.}
%Alternative: JWST-TST High contrast imaging I. New structures unveiled in the disk of Beta Pictoris

\author[0000-0002-4388-6417]{Isabel Rebollido}
\affiliation{Centro de Astrobiolog\'ia (CAB CSIC-INTA) ESAC Campus Camino Bajo del Castillo, s/n, Villanueva de la Cañada, 28692, Madrid, Spain}
\affiliation{Space Telescope Science Institute, 3700 San Martin Drive, Baltimore, MD 21218, USA}

\author{Christopher C. Stark}
\affiliation{NASA Goddard Space Flight Center, Greenbelt, MD 20771, USA}

\author[0000-0003-2769-0438]{Jens Kammerer}
\affiliation{European Southern Observatory, Karl-Schwarzschild-Straße 2, 85748 Garching, Germany}
\affiliation{Space Telescope Science Institute, 3700 San Martin Drive, Baltimore, MD 21218, USA}

\author[0000-0002-3191-8151]{Marshall D. Perrin}
\affiliation{Space Telescope Science Institute, 3700 San Martin Drive, Baltimore, MD 21218, USA}

% *** order of authors is TBD, for now I am just adding names into the LaTex to get this more complete --MP

\author{Kellen Lawson}
\affiliation{NASA Goddard Space Flight Center, Greenbelt, MD 20771, USA}

\author{Laurent Pueyo}
\affiliation{Space Telescope Science Institute, 3700 San Martin Drive, Baltimore, MD 21218, USA}

\author[0000-0002-8382-0447]{Christine Chen}
\affiliation{Space Telescope Science Institute, 3700 San Martin Drive, Baltimore, MD 21218, USA}

\author{Dean Hines}
\affiliation{Space Telescope Science Institute, 3700 San Martin Drive, Baltimore, MD 21218, USA}

\author[0000-0001-8627-0404]{Julien H. Girard}
\affiliation{Space Telescope Science Institute, 3700 San Martin Drive, Baltimore, MD 21218, USA}

\author[0000-0002-5885-5779]{Kadin Worthen}
\affiliation{William H. Miller III Department of Physics and Astronomy, John's Hopkins University, 3400 N. Charles Street, Baltimore, MD 21218, USA}

\author[0000-0002-7053-5495]{Carl Ingerbretsen}
\affiliation{William H. Miller III Department of Physics and Astronomy, John's Hopkins University, 3400 N. Charles Street, Baltimore, MD 21218, USA}

\author[0000-0002-8667-6428]{Sarah Betti}
\affiliation{Space Telescope Science Institute, 3700 San Martin Drive, Baltimore, MD 21218, USA}

\author[0000-0003-4003-8348]{Mark Clampin}
\affiliation{Astrophysics Division, Science Mission Directorate, NASA Headquarters, 300 E Street SW, Washington, DC 20546, USA}

\author[0009-0004-9728-3576]{David Golimowski}
\affiliation{Space Telescope Science Institute, 3700 San Martin Drive, Baltimore, MD 21218, USA}

\author[0000-0002-9803-8255]{Kielan Hoch}
\affiliation{Space Telescope Science Institute, 3700 San Martin Drive, Baltimore, MD 21218, USA}

\author[0000-0002-8507-1304]{Nikole K. Lewis}
\affiliation{Department of Astronomy and Carl Sagan Institute, Cornell University, 122 Sciences Drive, Ithaca, NY 14853, USA}

\author[0000-0001-9352-0248]{Cicero X. Lu}
\affiliation{Gemini Observatory/NSF's NOIRLab, 670 N. A'ohoku Place, Hilo, HI 96720, USA}

\author[0000-0001-7827-7825]{Roeland P. van der Marel}
\affiliation{Space Telescope Science Institute, 3700 San Martin Drive, Baltimore, MD 21218, USA}

\author[0000-0003-4203-9715]{Emily Rickman}
\affiliation{European Space Agency (ESA), ESA Office, Space Telescope Science Institute, 3700 San Martin Drive, Baltimore, MD 21218, USA}

\author[0000-0002-6892-6948]{Sara Seager}
\affiliation{Department of Earth, Atmospheric and Planetary Sciences, Massachusetts Institute of Technology, Cambridge, MA 02139, USA}
\affiliation{Department of Physics and Kavli Institute for Astrophysics and Space Research, Massachusetts Institute of Technology, Cambridge, MA 02139, USA}
\affiliation{Department of Aeronautics and Astronautics, MIT, 77 Massachusetts Avenue, Cambridge, MA 02139, USA}

\author[0000-0003-2753-2819]{R\'emi Soummer}
\affiliation{Space Telescope Science Institute, 3700 San Martin Drive, Baltimore, MD 21218, USA}

\author[0000-0003-3305-6281]{Jeff A. Valenti}
\affiliation{Space Telescope Science Institute, 3700 San Martin Drive, Baltimore, MD 21218, USA}

\author[0000-0002-4479-8291]{Kimberly Ward-Duong}
\affiliation{Department of Astronomy, Smith College, Northampton MA 01063, USA}

\author{C. Matt Mountain}
\affiliation{Association of Universities for Research in Astronomy, 1331 Pennsylvania Avenue NW Suite 1475, Washington, DC 20004, USA}  % Matt requested to be last author

%\affiliation{}https://www.overleaf.com/project/64230bcd88a9e2462ad30d31

\begin{abstract}
We present the first JWST MIRI and NIRCam observations of the prominent debris disk around Beta Pictoris. Coronagraphic observations in 8  filters spanning from 1.8 to 23~$\mu$m provide an unprecedentedly clear view of the disk at these wavelengths. The objectives of the observing program were to investigate the dust composition and distribution, and to investigate the presence of planets in the system. In this paper, we focus on the disk components, providing surface brightness measurements for all images and a detailed investigation of the asymmetries observed. A companion paper by Kammerer et al. will focus on the planets in this system using the same data. We report for the first time the presence of an extended secondary disk in thermal emission, with a curved extension bent away from the plane of the disk. This feature, which we refer to as the ``cat's tail", seems to be connected with the previously reported CO clump, mid-infrared asymmetry detected in the southwest side, and the warp observed in scattered light. We present a model of this secondary disk sporadically producing dust that broadly reproduces the morphology, flux, and color of the cat's tail, as well as other features observed in the disk, and suggests the secondary disk is composed largely of porous, organic refractory dust grains.

\end{abstract}


\section{Introduction} \label{sec:intro}

Beta Pictoris (\BPic) is perhaps the most iconic and well studied planetary system. Since the landmark discovery of \BPic's bright, edge-on debris disk \citep{Smith84}---making it the first exoplanetary system ever imaged---\BPic has been studied across the electromagnetic spectrum using both ground- and space-based observatories. Two decades ago, high angular resolution HST STIS coronagraphic observations revealed an inclined inner disk, the outer edge of which was used not only to argue for the presence of a putative perturbing companion, but also to constrain its mass and orbital distance \citep{Heap2000}. Since then, ground-based, near-infrared, high contrast imaging using VLT/NaCo has revealed the presence of \BPic b, a 10 M$_{Jup}$ companion with a semi-major axis of $\sim10$ au \citep{Lagrange09}, suggested to be consistent with the observed inner warp of the disk \citep{Dawson11}. More recently, long-term HARPS radial velocity monitoring revealed the presence of a second planet, \BPic c, interior to the first with a mass 9 M$_{Jup}$ and a semi-major axis of $\sim2.7$~au \citep{Lagrange19}, and subsequently confirmed by interferometry with VLTI/GRAVITY \citep{Nowak20}.

Multi-wavelength imaging is a powerful strategy for studying disks, as their appearance changes with wavelength, potentially revealing different components \citep[e.g.,][]{Hughes18}. Millimeter continuum images are primarily sensitive to thermal emission from large grains whose orbits are largely unaffected by radiation pressure and drag effects. As a result, ALMA images are believed to reveal the planetesimal birth rings. Visual observations, on the other hand, are primarily sensitive to scattered light from the smallest, sub-micron-sized grains. These grains are sensitive to radiation pressure and are expected to be blown out more efficiently around higher mass stars. As a result, HST STIS scattered light observations have revealed very extended disks, extending to hundreds of astronomical units from their parent planetesimal belts \citep{Schneider18}. Mid-infrared observations at $10-20~\mu$m are sensitive to thermal emission from warmer particles, and can reveal disk components that elude observation at either visible or millimeter wavelengths, especially with the great leap in sensitivity and resolution enabled by JWST. For instance, mid-infrared observations of Fomalhaut with JWST spatially resolved the warm inner disk and revealed an unexpected new intermediate dust belt, providing additional clues to dynamical processes within that complex system \citep{Gaspar23}. 

Detailed modeling of the history of our Solar System indicates that, over its lifetime, the dust production rate was dominated by individual collisions. However, whether individual collisions can affect the appearance of a debris disk is not well understood. Early modeling of debris disks assumed that the systems were in steady-state, constantly producing dust via collisions or outgassing and removing dust via radiation pressure and corpuscular stellar wind drag \citep[e.g.][]{StrubbeChiang06}. However, more recent Spitzer IRAC time monitoring of extreme debris disks (with optical depths $\tau = L_{\mathrm{IR}}/L_* > 0.01$) discovered that large collisional events in massive disks may be common \citep{Meng15}. Spitzer IRAC observations have also revealed fast and dramatic increases in the near to mid-infrared excess that have been attributed to collisional destruction of asteroid sized objects in the terrestrial planet zone \citep{Su22}. Since these collisions occur very close to the star, the dust they produce  is very challenging to spatially resolve. In parallel, high SNR HST scattered light imaging has discovered density enhancements in the more distant Kuiper Belt regions of debris disks \citep{Stark14,Jones23}. 

The dynamical outcome of a collision or outgassing event largely depends on the grain sizes produced.  Whether a grain is radiatively driven out of the system or gravitationally bound is determined by $\beta$, the ratio of the force due to radiation pressure compared with that due to gravity \citep{Artymowicz88}. Small grains tend to be more affected by radiation pressure and have larger values of $\beta$, while large grains can be relatively unaffected and have $\beta\sim0$. However, grains with the same $\beta$ may have a variety of different grain sizes, porosities, and compositions \citep{Arnold2019}. The near- and mid-infrared imaging enabled by JWST is expected to be broadly sensitive to grains near the blow out size limit ($\beta\sim0.5$) for typical debris disk host stars.

Several lines of evidence suggest that a significant dust production event may have recently occurred in the \BPic system. A clump of excess mid-IR emission detected on the southwest side of the disk \citep{Telesco05, Li12} has been interpreted as potentially due to production of dust from a massive collision of very large planetesimals ($\gtrsim 100$ km). Moreover, high contrast observations from the ground presented in \cite{Skaf23}  report a complex structure inwards of $\sim$ 60 au, with different clumps of material. Recent publications argued whether the main clump's position changes over time as if in Keplerian motion around the star \citep{Skaf23}, consistent with the recent collision hypothesis, or is instead stationary \citep{Han23}. Collisions and outgassing events should release gas as well, and indeed ALMA observations resolved a compact clump of gas spatially co-located with the mid-IR excess clump, detected first in $^{12}$CO \citep{Dent2014} and subsequently in \ion{C}{1} \citep{Cataldi18}.  Modeling of gas production and photodissociation rates for the clump, as well as the lack of azimuthal dispersion of the gas and dust, imply the production event was recent (within the last few thousand years or less; \citealp{Cataldi18}). 

Even a well-studied system such as \BPic is still capable of surprises. The Mid-Infrared Instrument (MIRI) coronagraphy presented in this work has revealed previously unseen additional components of the system: a sharply curved feature which extends up dramatically away from the disk plane, an extended asymmetric disk on the opposite side, and complex surrounding nebulosity. 
%Similarly, MIRI coronagraphy and 25 micron imaging also recently revealed new disk components in the Fomalhaut disk \cite{Gaspar23}. The fact that both of these disks, each already exceptionally well studied,  immediately revealed surprises in their first JWST observations is a testament to the dramatic leap forward in sensitivity at these wavelengths made possible with MIRI. 
We label the newly-seen curved feature of the \Bpic disk as the ``cat's tail'' based on its shape, and present below a dynamical hypothesis for its origin from radiation pressure acting on small particles in the aftermath of collisions. 

The remainder of this paper is organized as follows. Section \ref{sec:obs} presents the observations and data reduction processing, including high contrast Point Spread Function (PSF) subtractions and PSF deconvolutions. Section \ref{sec:morphology} presents results on disk morphology, describing the observed asymmetries and substructures seen in the disk, including the ``cat's tail'', followed by Section \ref{sec:disk_color} which presents results on the spatially resolved disk spectral energy distribution (SED) and colors. In Section \ref{sec:dynamics_model} we present a dynamical hypothesis to explain the origin of the ``cat's tail'' extended feature. We discuss the system comparing results presented here to previous observations in Sect. \ref{sec:discussion}; and we summarize and concude in section \ref{sec:conclusions}.

\subsection{Programmatic context}

Our collaboration is conducting several complementary studies to characterize the \BPic system with JWST.  A companion paper by Kammerer et al. (submitted) studies the planet \bpic b using the same NIRCam coronagraphy data presented herein. \bpic b is not visible in the MIRI coronagraphy data, because the thermal emission from the planet is hidden by the much greater thermal emission from the disk at 15 and 23~\um. Separate MIRI MRS observations of \Bpic (program 1294; PI C. Chen) have yielded thermal emission spectra of the disk, and also the planet at shorter mid-IR wavelengths where the disk is not yet so bright. These observations will be presented in Worthen et al.\ (submitted) and Chen et al.\ in prep. NIRSpec IFU observations for additional near-IR spectroscopic characterization of the disk were just obtained at the end of 2023 (program 1563; PI C. Chen). The characterization of the planetary system will be continued by NIRISS Aperture Masking Interferometry \citep[AMI,][]{sivaramakrishnan2023} observations (program 2297; PI T. Stolker), which aim to study the presence of clouds in the atmosphere of the inner planet ``c'' in particular. Collectively these programs are bringing a large fraction of JWST's full observing capabilities toward the detailed characterization of this key exoplanetary system.

This paper is part of a series to be presented by the JWST Telescope Scientist Team (JWST-TST)\footnote{\url{https://www.stsci.edu/~marel/jwsttelsciteam.html}}, which uses Guaranteed Time Observer (GTO) time awarded by NASA in 2003 (PI Matt Mountain) for studies in three different subject areas: (a) Transiting Exoplanet Spectroscopy (lead: N.  Lewis); (b) Exoplanet and Debris Disk High-Contrast Imaging (lead: M. Perrin); and (c) Local Group Proper Motion Science (lead: R. van der Marel). A common theme of these investigations is the desire to pursue and demonstrate science for the astronomical community at the limits of what is made possible by the exquisite optics and stability of JWST.
The first results in each of these areas are presented in \cite{Grant23}, \citep{Ruffio23} and \cite{Libralato23}, respectively.

%In addition to providing scientific support for observatory development through launch and commissioning, the team was awarded 210 hours of Guaranteed Time Observer (GTO) time. This time is being used for studies in three different subject areas: (a) Transiting Exoplanet Spectroscopy (lead: N. Lewis); (b) Exoplanet and Debris Disk High-Contrast Imaging (lead: M. Perrin); and (c) Local Group Proper Motion Science (lead: R. van der Marel). A common theme of these investigations is the desire to pursue and demonstrate science for the astronomical community at the limits of what is made possible by the exquisite optics and stability of JWST. 
The program on \BPic presented here was designed in particular to exercise JWST's coronagraphy modes over their full available wavelength range from NIRCam F182M to MIRI F2300C, on a key target that has played a central role in the study of exoplanetary systems since the field's inception.

\section{Observations and Data Reduction} \label{sec:obs}

We describe in turn the observations themselves, the reduction and calibrations of images individually and combined via PSF subtractions, and deconvolutions to reduce blurring of details by the JWST coronagraphs' off-axis PSF.  \BPic's disk is sufficiently bright to be immediately visible at very high SNR in all the observations, obvious in raw uncalibrated files before reduction or PSF subtractions. The disk is even spatially resolved in the MIRI target acquisition images, taken through a neutral density filter and with an exposure time of 2 seconds. Nonetheless we perform carefully optimized data reductions and image processing to maximize the visibility of fine details in the disk. 

\subsection{Observations}
\bpic was observed with MIRI \citep{MIRI,Wright2023}  and the Near Infrared Camera \citep[NIRCam;][]{NIRCam, Rieke2023}  on board JWST \citep{Gardner2023PASP..135f8001G}. Observations were performed as part of GTO program 1411 (PI C. Stark) on 2022 Dec 13 (MIRI) and 2023 March 18 (NIRCam)\footnote{The NIRCam observations were initially attempted on December 22, 2022, but failed due to a guiding issue. The observations were subsequently rescheduled roughly 3 months later, at a position angle rotated by approximately 90 degrees from the original plan.}. The observations used the coronagraphic imaging modes of both instruments \citep{Boccaletti22,Beichman2010,girard2022}. A summary of instrument settings is shown in Table \ref{tab:observations}.

Our NIRCam observations targeted \bpic in six different filters from $\sim 1.7$--$5~\um$, with filter coverage chosen to be sensitive to the presence of potential disk constituents such as water and CO ices. We used NIRCam's 210R coronagraphic mask for short wavelength (SW) observations; these were the first-ever coronagraphic science observations taken with SW channel. We used the 15 and 23 \micron\ coronagraphic filters for the MIRI observations, chosen for maximum sensitivity to thermal emission from the disk.

We adopted the common high contrast strategy of observing two roll angles on the science target plus a PSF reference calibrator. 
We observed the bright star $\alpha$ Pic as the PSF reference and included small grid dithers (SGD, \cite{2016SPIE.9904E..5KL}) for the MIRI 4QPM/F1550C and all NIRCam observations. SGD were deemed not necessary for the Lyot/F2300C combination, as the target acquisition accuracy was good enough, and the PSF is fairly extended. For MIRI, we obtained images of a nearby empty patch of sky to subtract astrophysical backgrounds as well as the ``glow-stick'' thermal background artifact discovered post-launch \citep{JWSTcomissioning}. 
To maximize the spatial extent of the disk visible, we positioned the disk's major axis at $\sim 45\deg$ to the 4QPM quadrant boundaries and the MIRI Lyot mask support strut using a position angle constraint. Similarly, we placed the disk diagonally across the NIRCam coronagraphic field of view.  For all filters we observed the disk at two different roll angles separated by $\sim10^{\circ}$. Observations were grouped in one non-interruptible sequence per instrument, to minimize the effects of temporal variations.

The total on-source exposure time was roughly one hour per each of the 8 filters used. We chose integration times and readout patterns to avoid saturation based on ETC calculations, which is particularly an issue at shorter wavelengths where the star is brighter. Integration times and total exposure duration on the brighter PSF calibrator were roughly half that on the science target.

\begin{table*}[t]
    \centering
    \begin{tabular}{c c l l l l l l }
         Instrument & Obs. dates & Coron.\ Mask/Filter & $\lambda$ & $\Delta\lambda$  &Readout Pattern & Exp. Time & Aperture PAs  \\
          &  &  & $(\micron)$ & $(\micron)$  & &  (s) & (degrees) \\
         
         \hline
         \multirow{6}{4em}{NIRCam}& \multirow{6}{8em}{2023 Mar. 18} & MASK210R/F182M & 1.84 & 0.24 & RAPID, 4, 90 & 1885.47 $\times2$ & \multirow{6}{6em}{84.3$\degr$, 94.3$\degr$}\\
         & & MASK210R/F210M & 2.09 & 0.21& RAPID, 4, 90 & 1885.47 $\times2$& \\
         & & MASK335R/F250M & 2.50 &  0.18 & BRIGHT2, 10, 80 & 1797.62 $\times2$& \\
         & & MASK335R/F300M & 3.00 & 0.32 & BRIGHT2, 10, 80 & 1797.62 $\times2$& \\
         & & MASK335R/F335M & 3.36 & 0.35 &  SHALLOW4, 10, 35 & 1871.53 $\times2$& \\
         & & MASK335R/F444W & 4.42 & 1.02 &  SHALLOW4, 10, 35 & 1871.53 $\times2$& \\
         \hline
        \multirow{2}{4em}{MIRI}& \multirow{2}{8em}{2023 Dec. 13} & 4QPM/F1550C & 15.50 & 0.90 &  FASTR1, 100, 55 & 1331.18 $\times2$& \multirow{2}{6em}{353.8$\degr$, 3.9$\degr$}\\
          & & Lyot/F2300C & 22.75 & 5.50 & FASTR1, 100, 55 & 1799.49 $\times2$& \\
         \\
    \end{tabular}
    \caption{Observing log for JWST observations of the \BPic disk in program 1411. The $\lambda$ column gives the pivot or center wavelength for each filter, and $\Delta\lambda$ gives the filter bandwidth. The readout column gives detector readout pattern, number of groups, and number of integrations. The exposure times listed are per roll, times two rolls as indicated for the total combined observing time on target. PSF calibrator and background observations were obtained with similar settings as the above.}
    \label{tab:observations}
\end{table*}

\subsection{Data reduction through stage 2}
\label{sec:data_reduction}

We downloaded the raw uncalibrated data (``uncal'' files) from the MAST archive and processed them with \texttt{spaceKLIP}\footnote{\url{https://github.com/kammerje/spaceKLIP}} \citep{kammerer2022,Carter23}. \texttt{SpaceKLIP} invokes the official \texttt{jwst} stage 1 \& 2 pipelines\footnote{\url{https://github.com/spacetelescope/jwst}} (used in its 1.12.1 version {\cite{pipeline_1_12_1}, with calibration files \texttt{CDRS\_CTX} $jwst\_1090.pmap$ and $jwst\_1088.pmap$ for MIRI and NIRCam respectively) with parameters optimized for coronagraphy (discussed in \citet{Carter23}), and provides further image processing and analysis steps to perform background subtraction, bad pixel cleaning, removal of 1/f noise, and PSF subtraction using either principle component analysis (PCA)-based or classical methods as described below. Achieving high-quality PSF subtractions depends critically on careful attention to pixel-level calibrations and bad pixel handling, and on image alignment, so substantial effort has been spent in optimizing these steps and their parameters.

For the MIRI data, as in \citet{Carter23}, we skipped the pipeline's dark subtraction step because the available darks for the coronagraph subarrays are of poor quality (low signal-to-noise ratio and a large number of bad pixels from e.g., cosmic rays). We set the pipeline's cosmic ray jump rejection threshold to 8, since this value was found to yield the best contrast performance by \citet{Carter23}, and adjusted the saturation correction step parameters so that only the left/right/upper/lower neighbors of a saturated pixel get flagged as bad pixels, as opposed to an entire 3 by 3 pixel box around a saturated pixel. In addition to these changes already implemented by \citet{Carter23}, we also skipped the pipeline's flat-fielding step because the available flats do not yet correctly describe the throughput loss due to the four-quadrant phase mask edges and the Lyot mask support, which created bright bar-like artifacts in the reduced images. By skipping the flat-fielding step, a flat-fielding error of typically less than 1\% for the Lyot mask and less than 2\% for the four-quadrant phase masks (as long as further than $\sim15$ pixels from the subarray edge) is introduced and propagated through our data reduction process. We discarded the first frame of each MIRI integration due to an increased level of noise caused by reset switch charge decay \citep{Carter23}. Then we identify and clean bad pixels as described in more detail in the next subsection. After bad pixel cleaning, we subtracted the corresponding background images from the science target and reference star images to remove the glow stick artifacts \citep{Boccaletti22,Carter23}. 

For the NIRCam data, we handled dark subtraction, jump rejection, and saturation correction similarly as for the MIRI data, but with a jump rejection threshold of 4 instead of 8. Moreover, we defined a custom border of four reference pixels around the edges of the NIRCam subarrays to remove offsets caused by detector bias drifts as discussed in \citet{Carter23}. \texttt{SpaceKLIP} now includes a custom pipeline step for correlated 1/f stripe noise removal on coronagraphic data.  It works on ramp data after jump detection by first making a preliminary estimate of the slope, subtracting that from each group to yield a residual image, fitting a Savitsky-Golay filter to the residual image to estimate the 1/f noise contribution in that group, and subtracting the noise model group-by-group prior to the ramp fit. Given the high SNR of this dataset, subtracting the 1/f noise yields a mostly cosmetic improvement in noise levels far from the star, but has negligible effect on the bright regions. Next, we median-subtracted each frame to remove any remaining background and cleaned bad pixels in a similar way as for the MIRI data.

\subsection{Bad pixel cleaning}

We paid particular attention to bad pixel cleaning.  We combined the DO\_NOT\_USE pixels from the pipeline's data quality extension with a custom bad pixel map of hot pixels. For MIRI, these additional hot pixels were identified by visual inspection of the MIRI background images. Through this procedure, we flag 28 additional hot pixels for the F1550C subarray and 48 ones for the F2300C subarray. For NIRCam, we first run the pipeline until and including the KLIP PSF subtraction step with the default DO\_NOT\_USE bad pixel map. Then, we inspect the PSF-subtracted frames by eye to identify additional hot pixels and re-run the entire pipeline, this time with an additional custom bad pixel map. For the short wavelength subarray, we flag no additional bad pixels, and for the long wavelength subarray, we flag 28 ones on top of the default DO\_NOT\_USE bad pixels. In both cases, all bad pixels are cleaned using a succession of three different algorithms. The first algorithm replaces pixels that are only bad in some (but not all) integrations with their median value in the good integrations. The second algorithm replaces remaining bad pixels with the median of their four direct neighbors (only considering the non-bad neighbors). The third algorithm replaces all remaining bad pixels with an image plane median filter of size five pixels using the \texttt{SciPy.ndimage.median\_filter} function.

\subsection{Image Alignments}\label{sec:alignment}

PSF subtraction depends sensitively on proper image registration of science and PSF reference observations. We recentered and aligned the NIRCam data as follows. For the first science frame of each dataset, we determined the true position of the host star \bpic behind the coronagraphic mask using phase cross-correlation with a simulated model PSF from \texttt{WebbPSF\_ext}\footnote{\url{https://github.com/JarronL/webbpsf_ext}} \citep{Leisenring2021,girard2022}. For this procedure, the bright core of the PSF was masked out as it is typically not represented well in simulations. Then the same Fourier shift and subtract least squares minimization was used to align all subsequent science and reference frames to the first science frame of each dataset. This alignment procedure has sufficient precision to recover the injected few-milliarcsecond small-grid dithers for the reference star and the $\sim1$~mas jitter ball from the pointing stability of \emph{JWST} \citep{rigby2023}.  As discussed further in the companion paper by Kammerer et al. (submitted), for this program the alignment analyses show nominal performance for the short wavelength observations (F182M \& F210M), but the long wavelength observations (F250M, F300M, F335M \& F444W) have an unusually large $\sim50$ mas pointing offset between the science and reference stars relative to the coronagraph mask. This slightly impacts the PSF subtractions at small separations, relevant to the planet \Bpic b, but has negligible impact on the disk seen at wider separations. 

We used the measured offsets to shift interpolate the PSF reference data to the science data prior to subtraction. To avoid ringing artifacts (i.e., Gibbs phenomena) from interpolating undersampled images, the F250M and F300M data were first convolved with a Gaussian filter. The filters had a full-width half maximum of 1.96~pixels and 1.71~pixels, respectively, chosen to  equal 125\% of the Nyquist sampling criterion of $\lambda/2.3D$, where $D = 5.2$~m is the effective pupil diameter in this mode behind the NIRCam round mask Lyot stop\footnote{\url{https://jwst-docs.stsci.edu/jwst-near-infrared-camera/nircam-instrumentation/nircam-coronagraphic-occulting-masks-and-lyot-stops}}.

For MIRI the situation was much simpler.  Following \citet{Carter23}, we simply assumed that the CRPIX position recorded in the FITS WCS header is the coronagraph mask center and also the star location for both science and reference targets, and aligned the images based on that. This simple assumption neglects the uncertainties in MIRI coronagraph target acquisition and in the precise knowledge of the exact mask center. These uncertainties are of order $\sim10$ mas, at least an order of magnitude less than the PSF FWHMs at these wavelengths. In any case the high brightness of the disk relative to the star at these wavelengths means that this simple approach suffices.

\subsection{PSF Subtraction strategies}

We tested and compared several methods for PSF subtractions, including classical and KLIP reference differential imaging (RDI), angular differential imaging (ADI), and model-constrained reference differential imaging (MCRDI). We ultimately found that for MIRI, classical RDI yields excellent results, while for NIRCam, the best disk images were achieved via MCRDI.

\begin{figure*}[p]
    \centering
    \includegraphics[width=\textwidth]{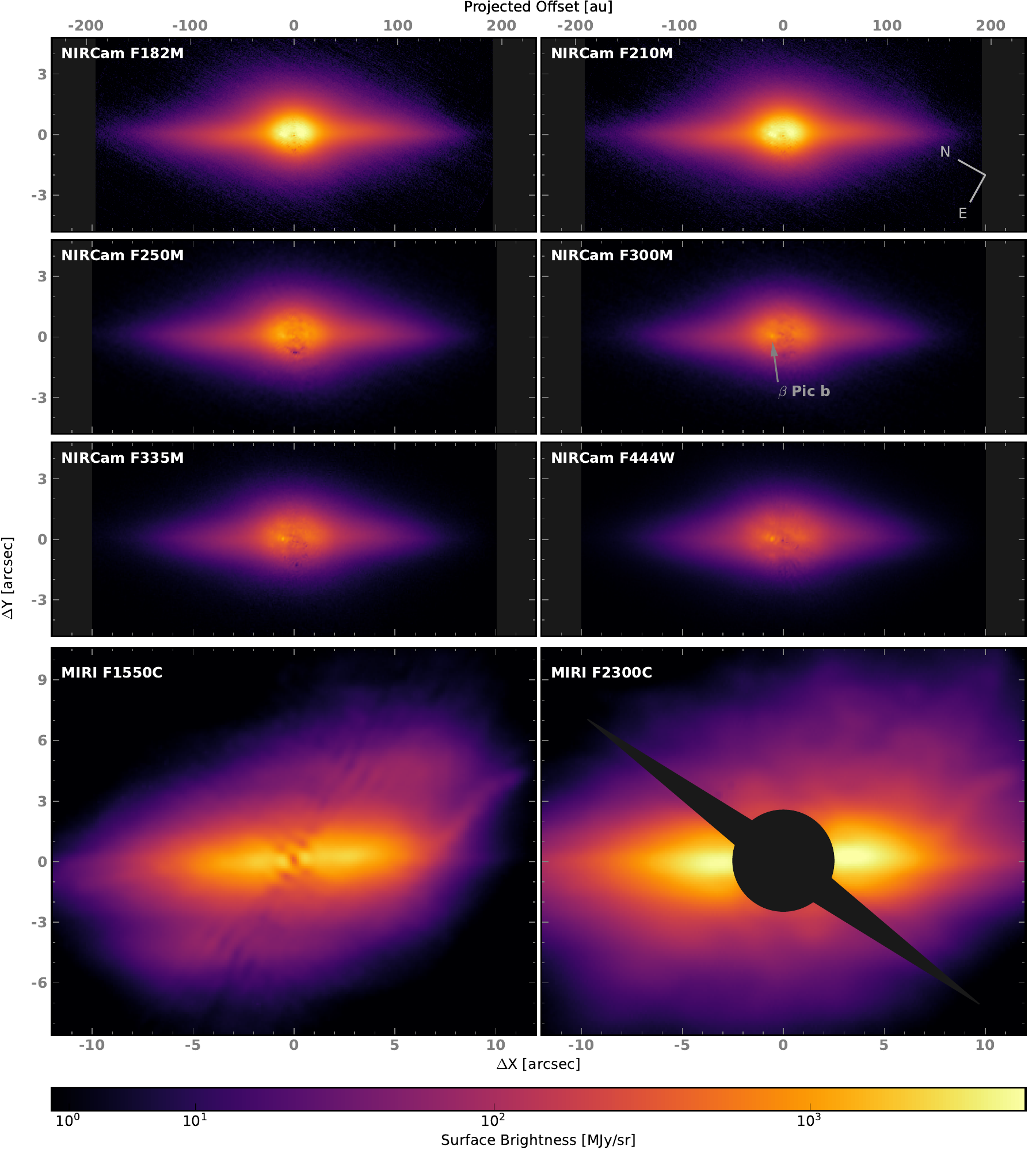}
    \caption{The \Bpic debris disk as seen with NIRCam and MIRI. The top 6 panels for NIRCam result from MCRDI PSF subtraction as described in section \ref{sec:mcrdi}, while the bottom two panels for MIRI are from classical PSF subtractions as described in section \ref{sec:classical_sub}.  In the NIRCam data, looking closely the planet \Bpic b can be seen just to the left (NE) of the occulted star (indicated with an arrow in the F300M data), along with the edge-on debris disk and its subtle warp already known from prior observations. In the MIRI data, a highly asymmetric curved feature we dub the ``cat's tail'' is seen to the west (SW), apparently bending sharply away from the disk plane (see Fig. \ref{fig:nebulosity_annotated} and Sect. \ref{sect:cats_tail}). A mask has been applied to the F2300C image only to block the pixels behind the opaque classical Lyot occulter and its supports. All images are shown rotated 60.9$^\circ$ to align the main disk major axis horizontally, on identical asinh stretches and identical spatial scales, though with a larger vertical field of view shown for the more extended nebulosity seen with MIRI.}
    \label{fig:psf_mcrdi_classical}
\end{figure*}

\subsubsection{Classical PSF Subtraction}
\label{sec:classical_sub}

For MIRI, the high brightness of the disk, relative faintness of the star, and extremely stable PSFs at these long wavelengths allow excellent results from straightforward classical PSF subtraction \citep{Krist97}. After alignment, we scaled the reference PSF images to match the observed stellar flux and then subtracted them. We scaled the reference images by minimizing the residuals in the PSF-subtracted images in regions where only speckle noise from the stellar PSF, but no significant disk flux was present. We then rotated the subtracted images from each roll to place north up. Finally, we averaged the images while weighting individual pixels by the coronagraphic mask throughput of the 4QPM for F1550C and the opaque Lyot occulter and its supports for F2300C, yielding the final images that are shown in the lower panels of Figure~\ref{fig:psf_mcrdi_classical}. Uncertainties after PSF subtraction were simply computed from those generated through the JWST pipeline, and first order propagations were applied accordingly.

We performed similar classical subtractions for the NIRCam data, but ultimately the MCRDI approach described in the next section was found to yield superior images.

\subsubsection{Model-Constrained Reference Differential Imaging}\label{sec:mcrdi}

Compared to classical RDI, techniques like KLIP \citep{Soummer2012} and LOCI \citep{Lafreniere2007} may better reproduce the stellar PSF's morphology by combining a set of reference images to minimize residuals with the science data. When circumstellar flux is present, such techniques necessarily overestimate the brightness of the starlight and therefore remove astrophysical signal as a side effect of subtracting the stellar PSF estimate from the data \citep[e.g.,][]{Pueyo2016, Ren2018}. This effect, known as ``oversubtraction", varies both spatially and spectrally \citep[e.g.,][]{Pueyo2012, Betti2022}, and thus introduces significant barriers for disk studies. Model Constrained RDI \citep[MCRDI;][]{Lawson2022} remedies this by suppressing the impact of circumstellar flux during construction of the stellar PSF model using a synthetic model of the circumstellar scene. 

To optimize a model of circumstellar signal (CSS) for use in MCRDI, a standard forward modeling procedure is used \citep[as in][]{Lawson2023}. In this approach, an initial (unconstrained) RDI reduction of the data is carried out, with each trial estimate of the CSS then being forward modeled and compared with these residuals to determine goodness-of-fit. For this purpose, optimization (of both the CSS and starlight models) considers only the region with stellocentric separations $0 \leq r < 2\farcs34$. PSF-subtraction is performed by determining the linear combination of the available reference frames that minimizes the (squared) residuals within this region. For all filters, the reference integrations at each dither position are median combined to yield a PSF library of five reference images.

We modeled the CSS as the superposition of a planet and a simple ring-like disk shown in Fig. \ref{fig:mcrdi_model}. To model the disk, we adopted a simplified version of the GRaTer code \citep{Augereau1999} as implemented in \texttt{Vortex Image Processing} \citep[\texttt{VIP};][]{Gonzalez2017} and assumed a scattering phase function (SPF) that is a linear combination of two Henyey-Greenstein (H-G) SPFs \citep{Henyey1941}. We varied disk parameters including inclination, position angle (PA), fiducial radius ($r_0$), scale height, the radial density power law indices interior and exterior to $r_0$, the disk flare, the vertical density exponent, the two H-G asymmetry parameters for the SPF, and the relative weight of the first H-G component. Following the procedure of \citet{Lawson2023}, for each roll angle of the data, we rotated a given raw disk model to the appropriate position angle, multiplied it by a coronagraphic transmission map, and then convolved it with a grid of (normalized) spatially-varying PSFs drawn from \texttt{WebbPSF-ext} \citep{Leisenring2021,girard2022}. Throughout this process, we accounted for the position of the coronagraph center in each exposure after alignment (see Section \ref{sec:alignment}). The grid of PSFs samples the origin as well as five logarithmically spaced radial positions at each of four linearly spaced azimuthal positions (for 21 PSF spatial samples in total). 

Nominally, the planet model component in the CSS would be governed by three parameters: two for position and a third for brightness. To simplify the optimization, we determined the planet's position \textit{a priori} using the F444W data. To do this, we first optimized the disk model while masking the region within $5 \lambda / D$ of the planet's estimated position. Subsequently, we tuned the planet's position by fitting the disk-model-subtracted residuals with PSFs drawn from \texttt{WebbPSF} \citep{Perrin2014}. The planet is then fixed to this position for all filters, reducing the planet model component to a function only of planet brightness. The overall CSS model is scaled post facto to minimize the residuals with the data. As such, the planet brightness parameter effectively governs the brightness of the planet relative to the disk.

We optimized the composite CSS model of 12 parameters separately for each NIRCam filter using the modified Powell algorithm as implemented in the \texttt{LMFit} package \citep{Newville2022}. For the short wavelength channel data only (F182M and F210M), we median combined the integrations for each exposure for the optimization process. This significantly reduces run time, as these filters' exposures have both more detector pixels and more integrations than the long wavelength filters. Once we identified an optimal CSS model, we performed MCRDI using that model to mitigate oversubtraction but otherwise using identical PSF subtraction settings as for the initial unconstrained reduction. Specifically, we construct the model of the stellar PSF by subtracting the optimized CSS model from the data when computing the optimal linear combination of the reference images. The resulting stellar PSF model is then subtracted from the original data. An example of the best-fit MCRDI model components is shown for the F335M data in Figure \ref{fig:mcrdi_model}.
In order to create an uncertainty map, we computed uncertainties for the stellar PSF model using standard error propagation for a linear combination of the reference images, and propagated uncertainties for subtraction of the PSF model from the science data. Finally, we used the standard uncertainty propagation for a median combination to get uncertainties for the final PSF-subtracted science images.

\begin{figure*}
    \centering
    \includegraphics[width=\textwidth]{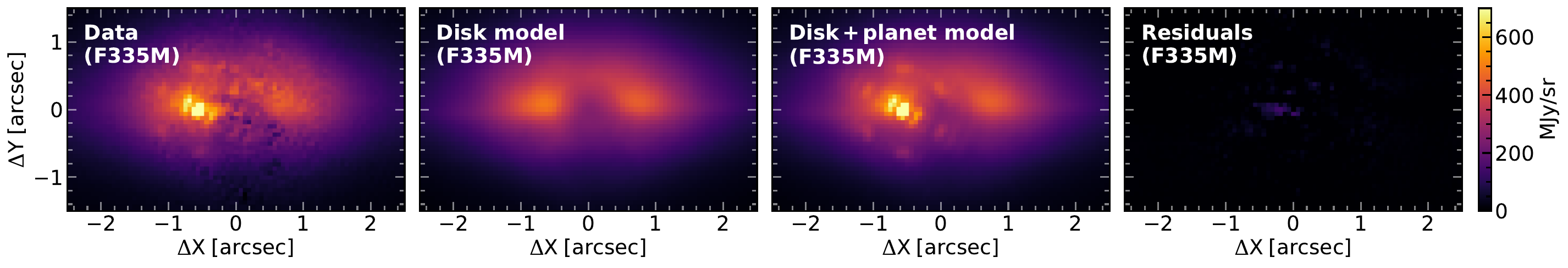}
    \caption{An example illustrating the components of the MCRDI model for the F335M data. Left: the reduced MCRDI image. Middle left: the best-fit PSF-convolved disk model. Middle right: the composite disk and planet model. Right: the final residual map for this model. The field of view is chosen to show the model within approximately the $r < 2\farcs34$ optimization region used for MCRDI. As in Fig. 1, the images were rotated by 60.9$^\circ$ to align the main disk with the horizontal axis.}
    \label{fig:mcrdi_model}
\end{figure*}

We note that the MCRDI process involves fitting models to the disk for use in avoiding oversubtraction, but we do not here attempt to interpret these as a rigorous or physically-correct model of the disk, as this model framework is intentionally simple.

For both the MCRDI and classical subtraction methods, the subtracted images from each roll angle are rotated and averaged together in the usual fashion. 
Figure \ref{fig:psf_mcrdi_classical} shows the \BPic system as seen in all filters after these PSF subtractions rotated 60.9$^\circ$ \citep[given a position angle of 29.1 $^\circ$ from ][]{Apai15} to align the main disk horizontally.

\subsection{Deconvolution}\label{sec:deconv}

We applied a deconvolution step to remove the blurring effect of the complex NIRCam and MIRI coronagraphic off-axis point spread functions, which arise due to the diffractive effects of the coronagraph Lyot stop pupil masks.  To deconvolve the data, we adopted the Richardson–Lucy (R–L) algorithm \citep{Richardson1972,Lucy1974} with a small alteration to accommodate coronagraphy. Specifically, for each iteration of the R–L algorithm, before convolving the prior iteration with the PSF, we multiplied it with a coronagraphic transmission map. Effectively, this includes the coronagraph's effects in the current estimate of the blurring at each step, providing better performance near the coronagraphic IWA. Though there are more recent methods designed to accommodate spatially varying PSFs \citep[e.g.,][]{Sakai2023}, we find that using a single PSF is sufficient for these NIRCam and MIRI Lyot data so long as the coronagraphic transmission is also considered as described. While the shape of the PSF for these data does change with detector position, the changes are sufficiently small that such an approach is not confounded; Appendix \ref{app:deconv_validation} demonstrates this through application to a synthetic disk model convolved with consideration for these spatial variations, where surface brightness measurements for the deconvolved result match those of the ground-truth (unconvolved) disk model to within $\lesssim1$\% beyond the IWA and $\sim 10$\% within the IWA. 

We note that for the MIRI F1550C data, the 4QPM introduces significant spatial variations in the PSF. Moreover, suitably precise coronagraphic transmission maps for the 4QPM are not yet available.  As such, our deconvolution approach is least well-suited for these data and induces significant artifacts, particularly at the edges of the field of view and the 4QPM boundaries. The deconvolved F1550C image of Figure \ref{fig:deconvolved} is therefore provided as a work-in-progress result and is not the sole basis for any morphological analysis herein.

Since bright circumstellar flux persists to the edges of the field of view for both MIRI filters, the deconvolution approach can introduce edge artifacts that increasingly infringe on the region of interest over successive iterations. To address this, prior to deconvolution we padded each initial MIRI image with a 35 pixel wide border of NaNs and then iteratively replaced the NaN values with the median of values within a $5\times5$ pixel window until none remained. Following deconvolution, this padded border is cropped from the final image. This approach is effective in preventing the aforementioned edge artifacts and induces no appreciable difference in the regions free of artifacts, though some small effect should be expected within a few $\lambda/D$ of the image edges.

Uncertainties for the deconvolved products are the same as those for the convolved images. 

Figure \ref{fig:deconvolved} shows the deconvolved disk images, on the same spatial scale and extent as Figure \ref{fig:psf_mcrdi_classical}. 

\begin{figure*}[p]
    \centering
    \includegraphics[width=\textwidth]{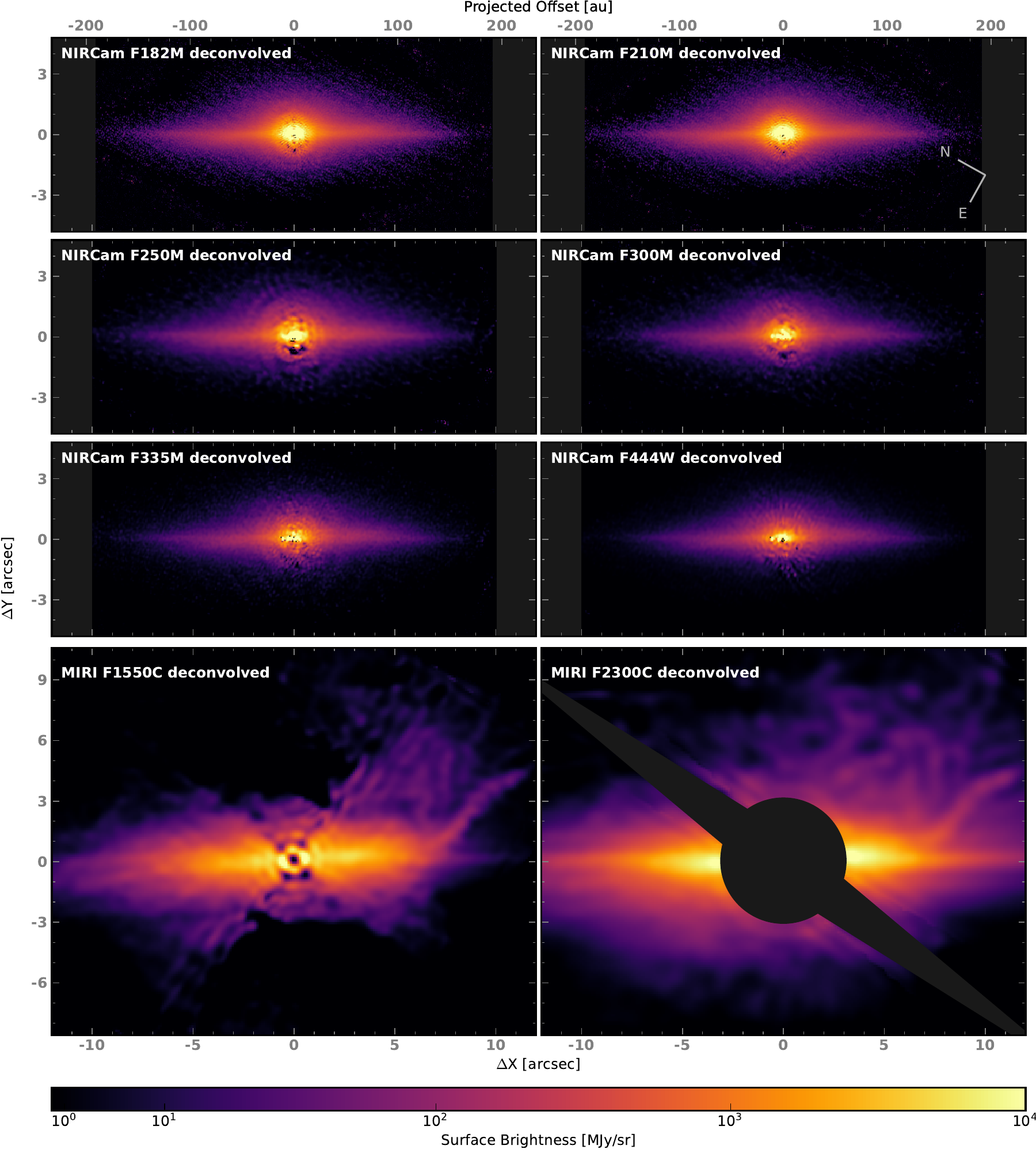}
    \caption{Same data as Figure 1 after deconvolution of the off-axis coronagraphic PSFs as described in section \ref{sec:deconv}. The presence of a vertical warp from the inclined secondary disk is seen more clearly in the deconvolved NIRCam data. Many complex asymmetric substructures are visible in the deconvolved MIRI data, consistently detected in both filters. Note that the deconvolution of the F1550C data is more challenging and thus has higher resulting uncertainties as discussed in the text. Also the masked region around the F2300C Lyot occulter and its supports has been expanded to block edge regions that are not reliably deconvolved. Similarly to Fig. 1, the images were rotated by 60.9$^\circ$ to align the main disk horizontally.}
    \label{fig:deconvolved}
\end{figure*}

\section{Disk Morphology} \label{sec:morphology}
%Figures \ref{fig:psf_mcrdi_classical} and \ref{fig:deconvolved} feature complex structures beyond a simple edge-on symmetric disk model. 

In this section we quantify the morphology of the observed asymmetries and surface brightness in order to facilitate comparisons with literature data and the model described in Sect. \ref{sec:dynamics_model}. Figure \ref{fig:nebulosity_annotated} shows the main features discussed in this section.

\subsection{Two misaligned disk components}
\label{sec:component_fits}

% How about: 
%

%Decades of past observations revealed a warp in the inner region ($\lesssim 4''$) of the disk \citep{Burrows95, Heap2000}, linked to a secondary disk misaligned with the main component and proposed to be originated via interactions with a possible planet \citep{Mouillet97,Augereau2001}. Hubble Space Telescope and ground-based adaptative optics (AO) coronagraphy enabled measurements of the secondary disk's appearance and asymmetries across the visible and near IR \citep[]{Golimowski06,Milli2014, Apai15}.
The JWST observations detect the presence of the secondary disk component, clearly observed in the MIRI images, both before and after deconvolution, and also present in the NIRCam images, though significantly fainter (see Figs.~\ref{fig:psf_mcrdi_classical} and~\ref{fig:deconvolved}). 
In the NIRCam data, the appearance of the secondary disk is broadly similar to prior Hubble \citep[e.g.][]{Heap2000, Golimowski06} and ground based \citep{Milli2014} observations: slightly inclined with respect to the main disk, superimposed on it within the innermost $\sim5$ \arcsec, and roughly symmetric between the northeast (NE) and southwest (SW) sides. In the MIRI data, particularly the F1550C, within the same few arcseconds the disk appears more vertically extended or thicker, plausibly due to thermal emission from the inclined secondary disk.

\begin{figure*}[t!]
    \centering
    \includegraphics[width=0.8\textwidth]{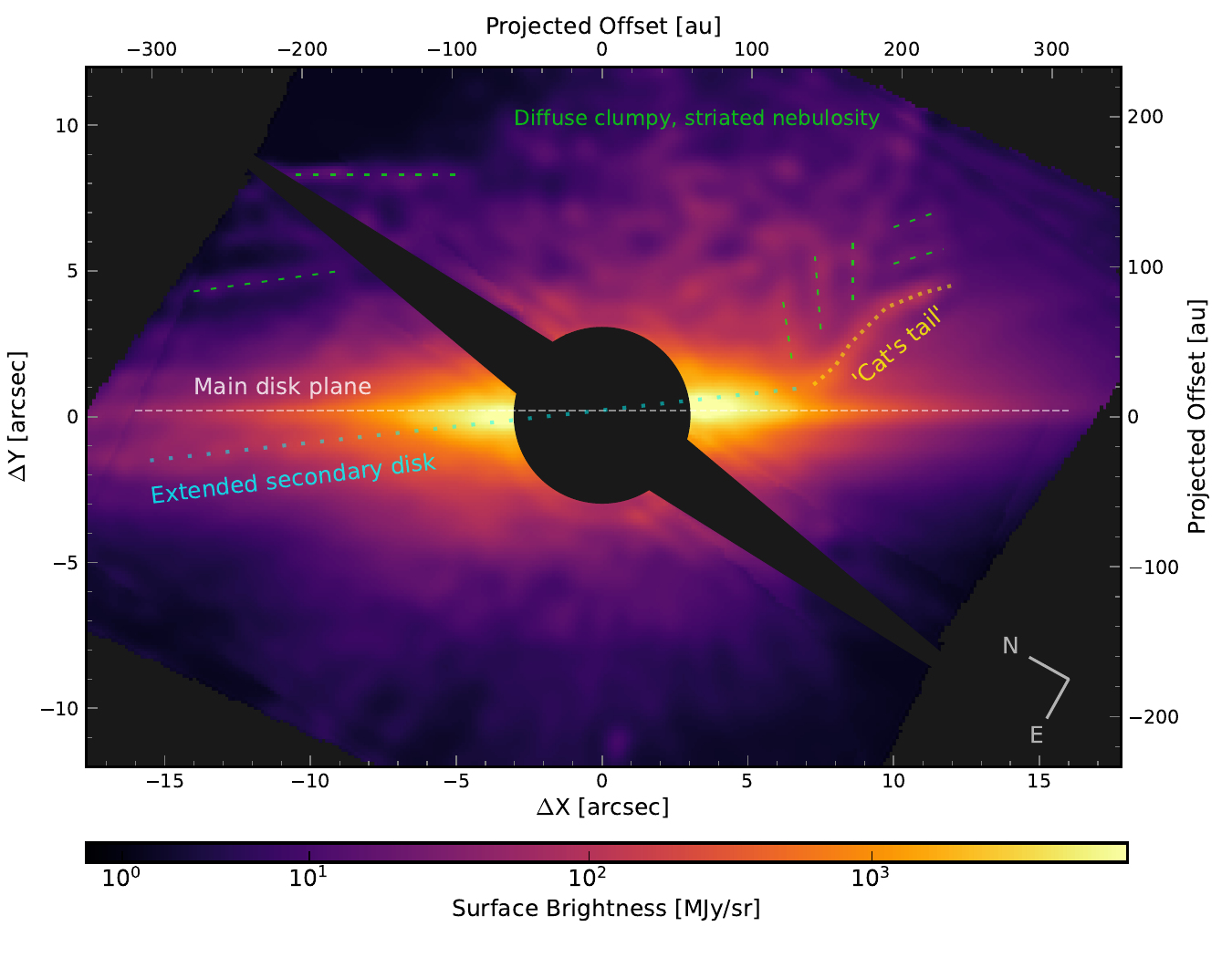}
    \caption{Complex nebulosity and substructures in the disk, as seen in F2300C. The stretch is adjusted to bring up fainter features observed in the outer regions. Relevant features are annotated including the main and secondary disk planes and cat's tail. The nebulosity to the west (top) includes both irregular clumps or knots, and also some apparent linear striations or tendrils. To the east (bottom) the clumps are fainter and fewer but still present, including one particular bright clump at bottom center. 
    These clumps are consistently detected in identical positions on the sky seen in the two distinct rolls, confirming they are astrophysical in origin rather than any instrumental artifact.}
    \label{fig:nebulosity_annotated}
\end{figure*}

At 15 and 23 \micron, the inner secondary disk appears to connect outwards to two distinctly different asymmetric features extending on either side. On the NE side, a long linear extension of the secondary disk extends out to at least 18\arcsec ($\sim$ 340 au, the edge of the MIRI Lyot field of view). This results in a bifurcated or forked appearance of the disk to the NE side, such as has been previously observed for a few other disks\footnote{For instance HD 111520; see figure 4 of \citet{Crotts2022ApJ...932...23C}.}.  On the SW side, instead we see a curved filament of material seemingly streaming off of the disk at a high angle ($\sim$45$^\circ$). On account of its raised, slightly S-curved appearance we refer to this feature as the ``cat's tail", described in more detail in the following subsection and Sect. \ref{sec:dynamics_model}.  

\subsubsection{Position angle of the main disk}
In order to better analyze the location and angle of the secondary disk, we performed a profile-fitting analysis similar to the one described in \cite{Golimowski06}. We investigated the projected vertical profile in the deconvolved images at different distances from the star, to extract a profile that traces both the main and secondary disks. 
First we measured the position angle (PA) of the main disk's midplane such that we can rotate the image and align the midplane of the main disk with the horizontal. 
We put the images through a high pass Gaussian filter with a 3 pixel sigma. We measured the average value of all the pixels in the high pass filtered image, and selected only those above with fluxes 10 times or larger than the average in order to include only high SNR points. We then fit a line to the main plane of the disk, obtaining slightly different PAs for each filter, listed in Table \ref{tab:SD_angles}. 
Uncertainties were obtained from error propagation of the reduced images and the line fitting process.

Given the significant offset between NIRCam and MIRI we investigated an alternative method to measure disk PA. To do so, we sequentially rotated the images over a range of possible PAs and fit a single Lorentzian component to vertical cuts in the disk. We then plotted the location of the peaks of the fitted function vs the distance to the star. That gives a line that we then use to validate our initial rotation angle vs the PA. If the rotation angle does \emph{not} correspond to the average disk PA, then the location of the peak of that best fit single component appears diagonal (say above the horizontal line on the NE and below on the SE). If the rotation angle does correspond to the average disk PA then the location of the single component peak is centered on a horizontal line ---albeit with some oscillation due to the presence of a second component, see below--- and appears symmetrical from a NE vs SW. More formally, the measured disk average PA is defined as the rotation angle that minimizes the absolute difference between the NE and SW locations of the peak of a single component model. 

Using this method we find results broadly consistent with Table \ref{tab:SD_angles}, and again indicating variation in PA with wavelength: $29.1$ $\pm$ $1.1 \degr$ for the NIRcam short wavelength channel,  $29.6$ $\pm$ $1.4 \degr$ for NIRcam long wavelength channel, and $30.4$ $\pm$ $1.5 \degr$ for MIRI filters.
%We calculated uncertainties based on the fitting process; such uncertainties are mostly statistical, since the instrumental error for the PA is negligible. 
The PA we measure in the NIRCam data is consistent with the value found by prior authors ($29.1 \pm 0.1\degr$ in \citealp{Apai15}, $29.0 \pm 0.2\degr$ in \citealp{Milli2014}), although we note the best fit PA found in $L'$ SPHERE observations in \cite{Milli2014} is 30.8$^\circ$. Literature reporting on  mid-IR observations \citep[e.g.][]{Li12,Telesco05,Han23}  suggest a PA $>$30$^\circ$. The MIRI data are consistent with that,  supporting the inference that the apparent major axis PA does vary with wavelength. At least a partial explanation to this can be found in \cite{Skaf23}, where they show in their Fig. 6 how the best PA varies for the two different sides of the disk. Besides the IR, the gaseous CO disk in \cite{Dent2014} was also inclined $\sim$ 4~$^\circ$ with respect to the dusty disk, observed in the continuum. 

Due to the variation in apparent major axis position angle between filters, for consistency of image display we decided to use the $29.1\degr$ angle obtained in the NIRCam short-wavelength channel analysis for the de-rotation of the images in all figures where we are depicting the major axis horizontal. Data analyses however used the individual best-fit position angles per each filter.

\subsubsection{The secondary disk}
After rotating the images to align the main disk with the horizontal, we binned the data in the horizontal direction to increase the SNR of vertical cuts. For the NIRCam data we adopted a horizontal bin width of four pixels while for the MIRI data we adopted a bin width of two pixels. We then performed a fit to each vertical cut using two Lorentzian curves simultaneously: the first one to match the position of the main disk, forcing the peak position to be within a few pixels of the main plane of the disk, and the second one free to trace the location of the secondary disk/fork. Results for the peak positions as a function of circumstellar distance for both components are shown in Appendix \ref{app:secondary_disk}, Figs.\ref{fig:SD_nircam} \& \ref{fig:SD_miri}. 

We performed a fit for the secondary disk in both NIRCam and MIRI, including ranges from -100 to 100~au, where the disk is detected. 
The angle of the secondary disk with respect to the main disk for the different filters are summarized in Table \ref{tab:SD_angles}.
 
The inner secondary disk seen using MIRI connects with the asymmetric extension creating a ``fork" on the NE side, and the ``cat's tail" on the SW side, as shown by Fig.\ref{fig:nebulosity_annotated}~and~\ref{fig:SD_miri}. In order to further characterize these observed features we extracted their inclination angles with respect to the main disk. We distinguished two different ranges in the SW side for the MIRI images: (1) base of the cat's tail, from 120~au until 190~au; and (2) tip of the cat's tail, from 190~au until 230~au. The angles with respect to the main disk of the two regions can be found in Table \ref{tab:SD_angles}.
%The secondary disk shows a small variation in inclination, specially in the SW side, possibly due to the presence of structures on the west side. 

\begin{table}[]
    \centering
    \begin{tabular}{lccccc}
     & Main disk$^*$ & \multicolumn{4}{c}{Components of secondary disk$^{\dagger}$} \\
    \hline
        Filter & PA & Sec. & NE & Cat's tail & Cat's tail \\
               &     &  disk  &   fork & base & tip \\
            & (deg) & (deg) & (deg) & (deg) & (deg) \\
        \hline
        F182M& 29.6$\pm$ 1.5& 5.8 & -- & -- & -- \\
        F210M& 29.6$\pm$ 1.1& 5.9 & -- & -- & -- \\
        F250M& 28.0$\pm$ 1.2& 6.2 & -- & -- & -- \\
        F300M& 28.0$\pm$ 1.9& 5.9 & -- & -- & -- \\
        F335M& 28.0$\pm$ 1.9& 5.0 & -- & -- & -- \\
        F444W& 28.9$\pm$ 1.7& 5.1 & -- & -- & -- \\
        \hline
        F1550C& 31.6 $\pm$ 1.0& 5.9 & 5.5 & 45.0 & 21.5 \\
        F2300C& 33.2 $\pm$ 1.9& 5.6 & 4.8 & 44.7 & 21.1 \\
        
    \end{tabular}
    \caption{PA of the main disk, and angles measured for the different components of the secondary disk. $^*$PA angles of the main disk refer to the counterclockwise angle measured from the north vector. $^{\dagger}$ Angles of the components of the secondary disk refer to the relative location with respect to the main disk. }
    \label{tab:SD_angles}
\end{table}

Overall, the secondary disk is prevalent in all images on the NE side, but harder to detect in the SW side. We also see a dependence with wavelength in the NIRCam images, where the secondary disk becomes fainter at longer wavelengths, indicating the grains are efficient at scattering light at visual and short NIR wavelengths. As the thermal emission starts to dominate, the secondary disk then picks up again in brightness at MIRI wavelengths, as well as becoming more spatially extended to the NE creating the appearance of a fork.

\subsection{The cat's tail in $\beta$ Pic}
\label{sect:cats_tail}

The most conspicuous feature observed for the first time in this JWST dataset is the prominent curved structure on the SW side of the disk, resembling a ``cat's tail", seemingly attached to the disk. We observe it in both MIRI filters, at 15.5~$\mu$m and 23~$\mu$m, and it is present in both rolls.

The structure extends from $\sim$ 6.5~\arcsec (from the location of the star) to $\sim$ 13.5~\arcsec, corresponding to $\sim$125 to 265 au. Details of how the structure were characterized are in the previous section, Table \ref{tab:SD_angles} and Fig. \ref{fig:SD_miri}. The base of the cat's tail separates from the disk near a projected separation of $\sim6$ \arcsec (115 au) and extends for $\sim3$~\arcsec (60 au), curving away from the main disk and reaching a position angle of $\sim$45~$^\circ$ with respect to the main disk midplane (see Table \ref{tab:SD_angles}). The tip of the cat's tail is considerably shorter, $\sim$1.5~\arcsec (30 au), and shows a very similar inclination in both filters of $\sim$21$^\circ$.

The cat's tail has not been reported before, despite deep HST observations with several instruments \citep[see e.g.][]{Burrows95,  Golimowski06, Apai15, Ballering16}. There is also no evidence for such structure in ALMA data \citep{Dent2014, Matra19}, nor any of the ground-based mid-IR observations \citep[e.g.][]{Telesco05,Li12, Skaf23}. There may be faint hints of material near this location in the outermost part of the $L'$ band imaging presented in \cite{Milli2014}, but only at the edge of the FOV where PSF subtraction residuals make any potential detection highly uncertain.  Nor do the JWST NIRCam observations show any sign of the cat's tail,  setting $3 \sigma$ upper limits of 0.23 and 0.01 mJy/arcsec$^2$ in F210M and F444W respectively for any excess emission at the location of the center of the tail.

In the upcoming sections we analyze the color (Sect. \ref{sec:disk_color}) and possible origin (Sect. \ref{sec:dynamics_model}) of the cat's tail. Additionally, further MIRI observations have already been performed at the time of submission of this paper for the program GTO 1241 (PI Choquet), including the filters F1140C and F1065C. These observations will help unveil more details about the nature of the cat's tail. 

\subsection{The nebulosity in the west side of the disk}

The MIRI F1550C and F2300C images show in the west region of the FOV (above the midplane of the main disk as shown in Figs. \ref{fig:psf_mcrdi_classical} and  \ref{fig:deconvolved}) faint small structures that we will refer to as ``nebulosity" from here on. The eastern region of the image (below the midplane of the main disk), also shows a few fainter such features. 
Figure~\ref{fig:nebulosity_annotated} shows the F2300C deconvolved image with relevant features annotated and a stretch that favors the observation of the nebulosity. These structures are not observed to have any symmetry axis with respect to the main disk or the star, and apparently, they do not have a preferential orientation. Some of this nebulosity has a striated or tendrilous appearance, in particular a few parallel linear clumps above the ``cat's tail" region, and a few rough lines to the north and not quite parallel with the disk plane. 

We confirm these features are \textit{not} instrumental artifacts or residual stellar speckles. These features are observed consistently in both rolls for both MIRI filter observations (i.e., four independent exposures, using two completely different sets of coronagraph optics within MIRI).  The fact that these features are fixed to the plane of the sky, independent of observatory position angle, unambiguously confirms they are astrophysical in origin. Also there is no sign of such nebulosity present in the reference star images.  In Sect.~\ref{sec:dynamics_model} we provide a hypothesis of its possible origin as part of the \bpic system, and its connection with the secondary disk and the ``cat's tail".

We cannot yet strictly rule out that some portion of these clumps may be mid-IR background sources in chance alignment; however the absence of any shorter-wavelength counterparts in the HST~\citep[e.g.][]{Golimowski06,Apai15} and NIRCam data rules out most classes of potential astrophysical background object. The proximity to \Bpic of so many background sources, all of which are so extremely red as to be only visible at $\geq 15$ microns, would be very low probability. Further the apparent alignment of the few linear streaks near the cat's tail and others roughly aligned to the disk plane suggests physical association of some sort.  We anticipate that proper motion analyses using multiple epochs of MIRI observations will conclusively rule out background status for all or nearly all of this nebulosity. 
%Whether they are linked to the \bpic disk or are background sources will require further investigation. 

\subsection{Surface brightness}
\label{sec:subsec_surfacebrightness}

Measurements of the surface brightness of the disk can reveal asymmetries in the dust distribution, and provide an overall view of the shape and flux of the disk with respect to the position of the star.

In order to investigate the surface brightness, we used the deconvolved images shown in Fig. \ref{fig:deconvolved} with the main disk major axis rotated horizontal according to the measured PA.
%We rotate the disk by the average PA from Table \ref{tab:SD_angles} for each instrument.
We perform an integration in the vertical direction (i.e., perpendicular to the main disk) without binning. 
We obtained values for the disk surface brightness along the whole FOV with the exception of the location of the star. The IWA varies for the different instruments and modes, according to the technical documentation \footnote{See technical details in:  \url{https://jwst-docs.stsci.edu/jwst-mid-infrared-instrument/miri-observing-modes/miri-coronagraphic-imaging} and \url{https://jwst-docs.stsci.edu/jwst-near-infrared-camera/nircam-observing-modes/nircam-coronagraphic-imaging}}. 
This is reflected in the measured SNR when extracting the profiles. We chose to only report measurements for which SNR$>$15 for all datasets. As a result, for NIRCam we can go down to 0.2\arcsec~ and still retrieve some structure, while for MIRI the IWA goes up to 2.7\arcsec~for the Lyot stop at 23~$\mu$m (4.6$\lambda$/D), in agreement with the expected performance \citep{Boccaletti2015,Boccaletti22}. For the 4QPM at 15.5~\um we excluded a larger portion of the inner disk, due to the PSF artifacts present even after deconvolution, leaving an effective IWA of 1\arcsec, instead of ~0.5\arcsec. 
The cutoff limit for the detection of the disk's outer edge was set to SNR$>$5.

We present the surface brightness for both the NE and SW sides, the contrast, and contrast adjusted for the projected stellocentric distance in Fig. \ref{fig:SB_CC_JWST}. 

The observed surface brightness decreases with wavelength for the NIRCam filters, as the grains become less effective to scatter the stellar light. But as the thermal emission picks up in the MIRI filters, the surface brightness of the disk starts to increase with wavelength (top left and right panels). Contrast levels also increase significantly with wavelength, as the star becomes less bright middle panels; and when corrected for the effect in flux of projected stellocentric distance (r$^2$), the contrast remains constant for NIRCam data from 30 up to $\sim$150 au, where the main disk becomes fainter bottom panels. The distance corrected contrast for MIRI, however, shows an abrupt decrease after $\sim$100~au in the SW side (bottom left panel), specially for the 15.5\um filter. This can also be observed in the images (e.g. Fig.\ref{fig:deconvolved}), as the cat's tail picks up.  
It's worth noting as well, that the surface brightness profiles for NIRCam filters show a small excess in the NE side at $\sim$10~au likely due to the presence of \bpic b.

\begin{figure*}
    \centering
    \includegraphics[width=.49\textwidth]{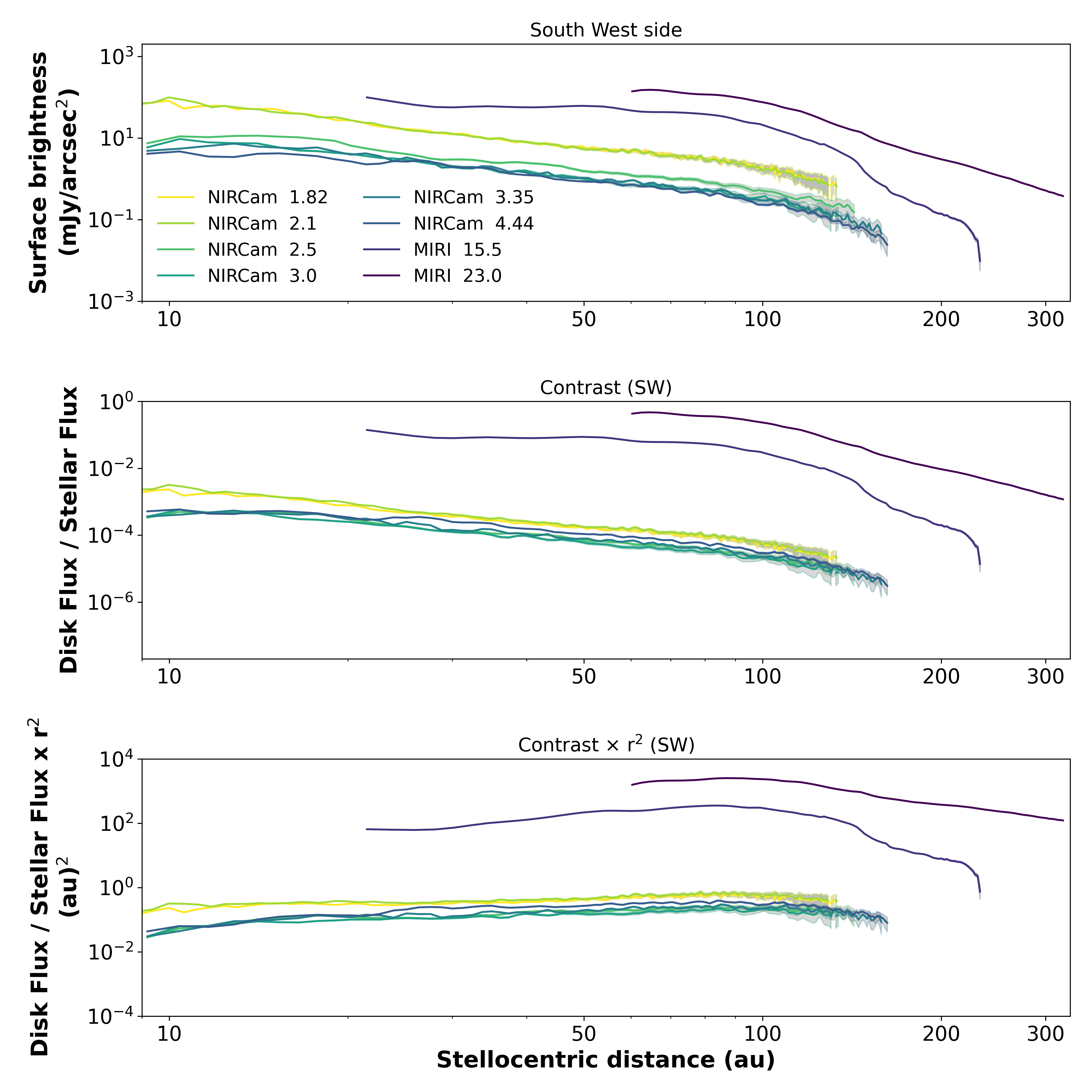}
    \includegraphics[width=.49\textwidth]{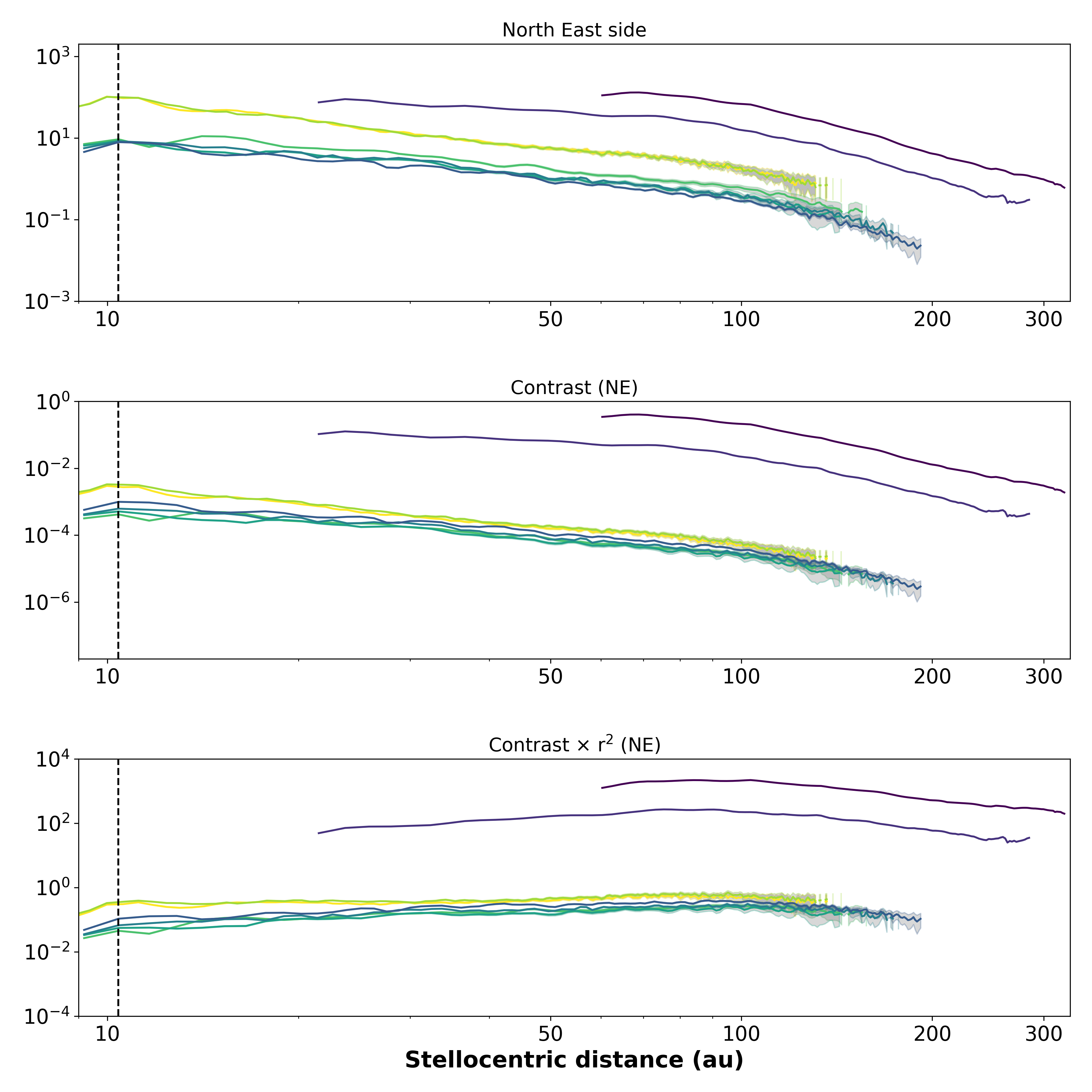}
    \caption{Surface brightness vertically integrated profiles of the disk in the different JWST filters, as labeled. Only regions with SNR$>$15 were investigated. Top panels show the surface brightness versus stellocentric distance to the star for the SW and NE side left and right,respectively. Middle and bottom panels show the contrast to the stellar flux and the contrast corrected for the projected distance dependence for the SW (left) and NE(right) sides. Shaded grey areas correspond to 3$\sigma$ range, color coded for each filter. The vertical dashed line in second panel (NE) indicates the location of \bpic b.}
    \label{fig:SB_CC_JWST}
\end{figure*}

When comparing the surface brightness of both sides of the disk, we find that the surface brightness levels along the disk are similar, yet the flux in the SW side decays faster than in the NE side after $\sim$100~au in all filters. This shorter extension of the SW side is well established in the literature \citep[see e.g. Fig. 8 in][]{Janson21}. There is no clear evidence of large asymmetries or brightness peaks/clumps in the NIRCam data. 

However in both MIRI filters there is significant spatial variation between NE and SW sides, and presence of localized peaks or clumps in brightness  between $\sim$40 and $\sim$100~au, shown in Figs. \ref{fig:asymmetry_JWST} and \ref{fig:miri_linear_r2}. This asymmetry, with larger flux in the SW side of the disk, seems to be co-located with the CO and dust clumps previously reported in the literature \citep[e.g.][]{Telesco05,Dent2014,Cataldi18}. We observe some structure in the asymmetry, with one brighter peak, close to 50~au on the SW side, and at least a second peak at $\sim$80~au with a significantly lower brightness. 
In order to further test the connection of the observed asymmetry to previous observations, we estimate the location of the brightest ``clump" through a simple Gaussian fit (see Fig. \ref{fig:asymmetry_JWST}, dashed lines). We obtain peak locations at 55.5$\pm$3.9~au for F1550C, and 63.4$\pm$2.9~au for F2300C, where uncertainties are from the fit results using the per-pixel uncertainty maps from the data reduction process as weights. %Both measured distances are significantly larger than previous estimations \citep{Telesco05,Skaf23,Han23}, but are still within the CO clump size detected with ALMA \citep{Dent2014}.
In the F2300C data, the observed peak location is not far outside of the area blocked by the large Lyot coronagraph occulting spot; the apparent difference in peak location between the two filters may arise at least in part from instrumental systematics. 
In order to compare with previous measurements in the literature, we include in Fig. \ref{fig:asymmetry_JWST} the central locations measured in \cite{Skaf23} and \cite{Han23}. Fig. \ref{fig:miri_linear_r2} top panel compares the MIRI F1550C data to the locations of the four clumps C1-C4 reported by \citet{Skaf23} based on 12 \micron\ observations from the ground (their C1 corresponds to the main brightness clump as detected by several prior works). The C1 and C2 clumps are clearly detected in the MIRI 15 \micron\ data. The other two are at sufficiently smaller separations  that PSF subtraction and deconvolution residuals make identification difficult. 
The presence and nature of the mid-infrared brightness asymmetries is further discussed in Sect. \ref{sec:discussion}.

\begin{figure*}
    \centering
    \includegraphics[width=0.8\textwidth]{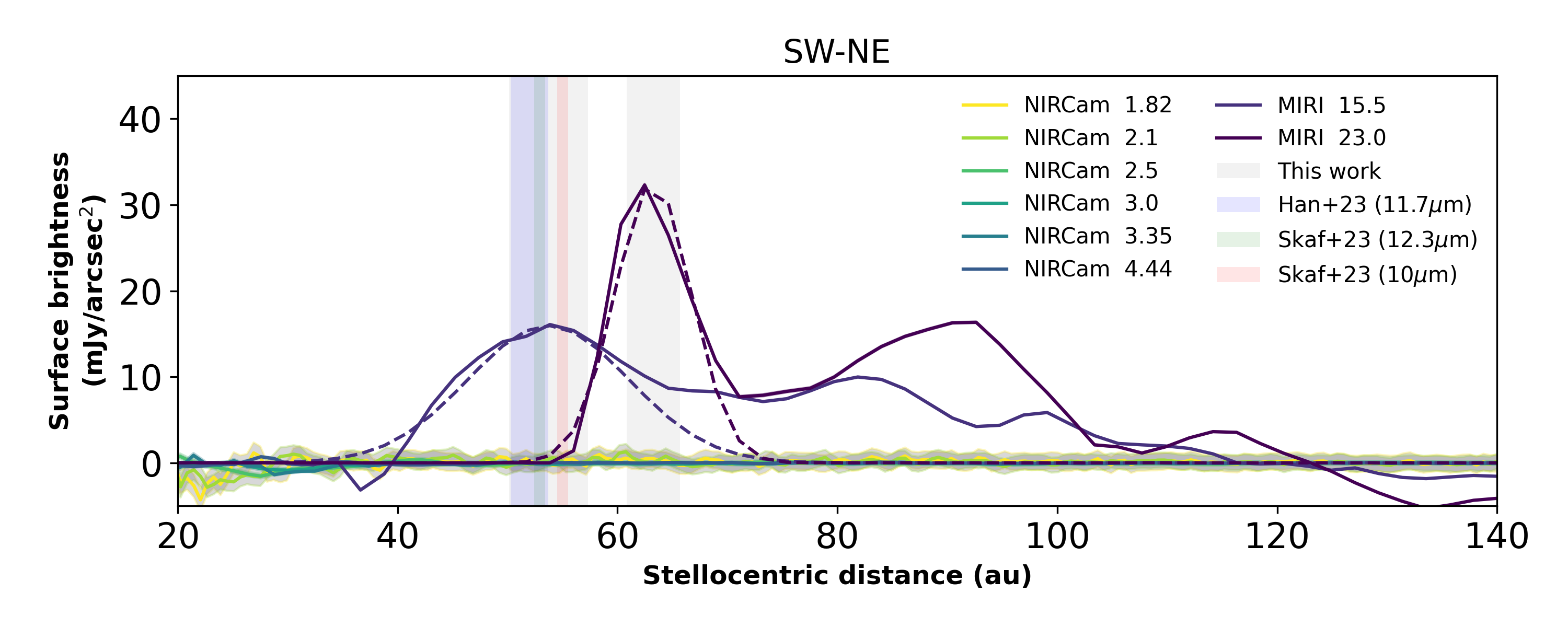}
    \caption{Difference between the surface brightness of the SW and NE sides of the disk. Asymmetric peaks or clumps are detected with MIRI between 40 and 100~au, with both the peak positions and brightnesses varying significantly between 15 and 23~\um. Dashed lines show the Gaussian fit used to estimate the central location of the brightest peak in the asymmetry at each wavelength. Shaded grey areas correspond to 3$\sigma$ ranges.  Color coded shaded areas correspond to results from prior observations at 10--12~\um. No significant variation between SW and NE sides is seen in the NIRCam data for this range of separations.}
    \label{fig:asymmetry_JWST}
\end{figure*}

\begin{figure*}
    \centering
    \includegraphics[width=\textwidth]{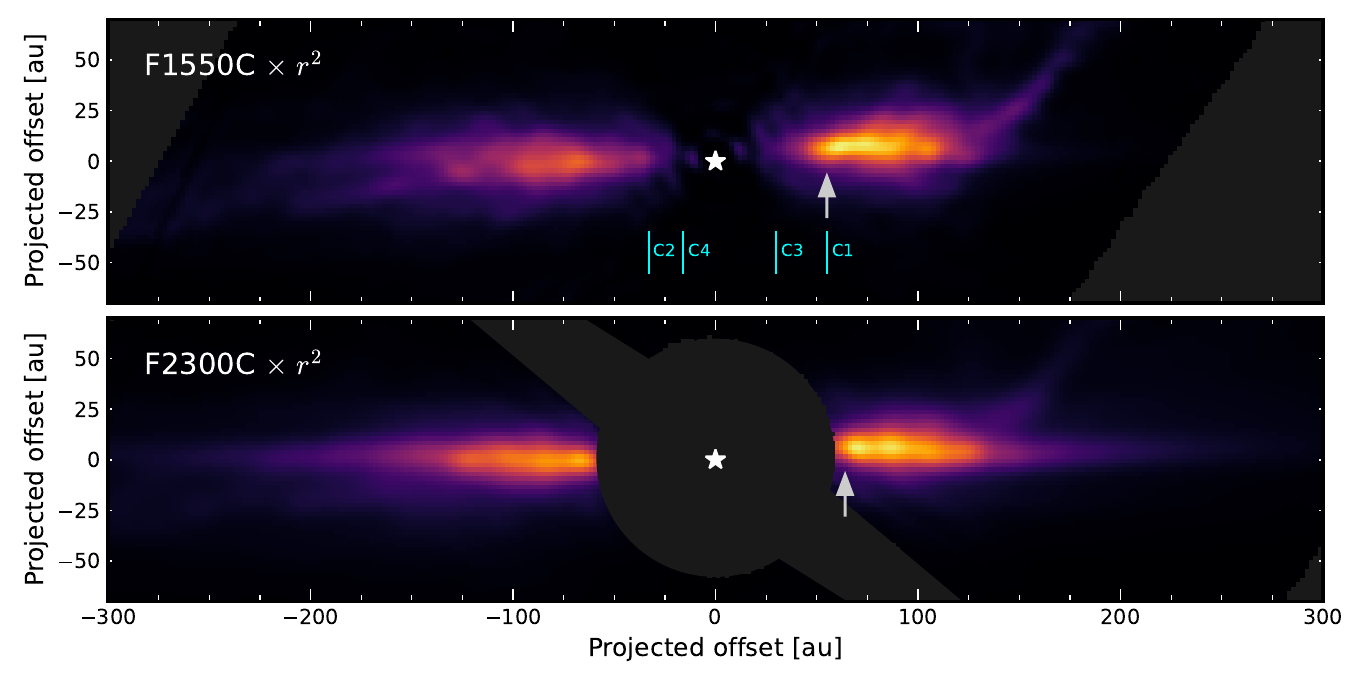}
    \caption{The MIRI coronagraphy data in F1550C and F2300C (15.5 and 23 \micron) presented with a linear stretch and scaled by the projected distance from the star squared. This stretch better visually emphasizes (a) the brightness asymmetry between the two sides, and (b) the presence of localized peaks or clumps of brightness particularly to the SW. The white arrows indicate the estimated clump locations from the Gaussian fitting shown in Fig. \ref{fig:asymmetry_JWST}, at 55.5$\pm$3.9~au for F1550C, and 63.4$\pm$2.9~au for F2300C. The cyan annotations C1-C4 indicate the angular separations of the clumps reported in ground-based VLT/NEAR 12 \micron\ data \citep{Skaf23}. Both C1 and C2 are clearly detected in the MIRI F1550C data. }
    \label{fig:miri_linear_r2}
\end{figure*}

\section{Disk Spectral Energy Distribution and Colors}
\label{sec:disk_color}
Spectral coverage across the near- and mid-IR enables color measurements that can help place constraints on the physical properties of disk constituents. In this section we explore the differences in the colors along the disk.

The main disk of \bpic does not present any conspicuous color differences between the NE and SW sides with the notable exception of the difference in the location of the dust clump (Sect. \ref{sec:morphology} and Fig. \ref{fig:asymmetry_JWST}). However, if we compare the secondary disk, particularly in the MIRI filters, we see that it is significantly bluer than the main disk. Figure \ref{fig:false_color} combines F1550C and F2300C data in a false color image, in which the contribution of the secondary disk stands out as bluer. Even though some instrumental effects remain in the F1550C image (annotated in the figure), the nebulosity and the cat's tail also exhibit a much bluer color than the main disk, possibly indicating a higher temperature of the grains. 
Figure \ref{fig:catstail_MD} shows the peak flux at different stellocentric distances for the cat's tail. In order to avoid contamination from the main disk, we started the measurement of the cat's tail at distances larger than 5.2\arcsec ($\sim$ 100 au), where the vertical separation between peak flux of the cat's tail and the main disk is $\sim$1~\arcsec at 15.5\um. We note that some contamination from the vertical component of the main disk might still be present, but removing it completely would require precise and extensive modeling, which is out of the scope of this paper. The left panel of Fig. \ref{fig:catstail_MD} shows the ratio of the peak flux of cat's tail to the main disk; and the right panel shows the absolute peak flux. The beginning of the cat's tail is marked in both panels with a red shaded area. The cat's tail is much brighter with respect to the main disk at 15~\um, and seems to increase with distance until peaking at $\sim$210~au, then decreasing until the detection is no longer significant. In contrast at 23~\um  the cat's tail is fainter than the main disk, and the the ratio is approximately constant with separation. In the right panel, an increase in flux is observed in both filters around $\sim$130~au. Beyond that, both filters show a different behaviour of the flux with distance, with an exponential decrease in F1550C and a quadratic decrease in F2300C. Interestingly, there is an increase both in total flux (right panel) and contrast (left panel) at $\sim$200~au, just after the change in the orientation angle, which may indicate there is some internal structure in the cat's tail itself. 

\begin{figure*}
    \centering
    \includegraphics[width=\textwidth]{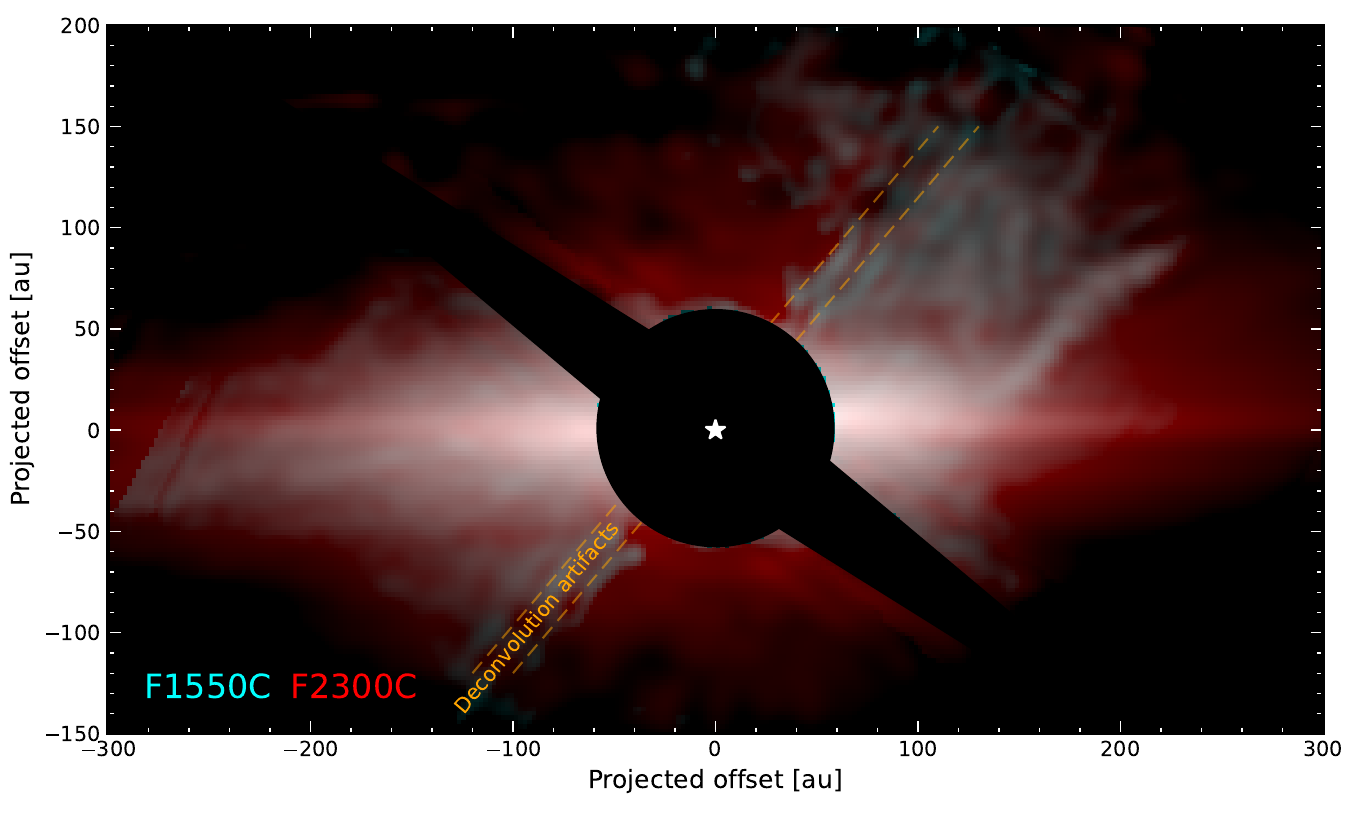}
    \caption{False color image generated from the F1550C (15.5~\um, cyan) and F2300C (23~\um, red) images. At these wavelengths, the main disk is redder, while the secondary disk, cat's tail, and extended nebulosity are bluer. The several linear striations or tendrils just above the cat's tail are clearly seen, and consistently blue. As noted in the text, the deconvolution of the F1550C is challenging and leaves residual artifacts along the 4QPM boundaries, as marked.
    }
    \label{fig:false_color}
\end{figure*}

\begin{figure*}
    \centering
    \includegraphics[width=\textwidth]{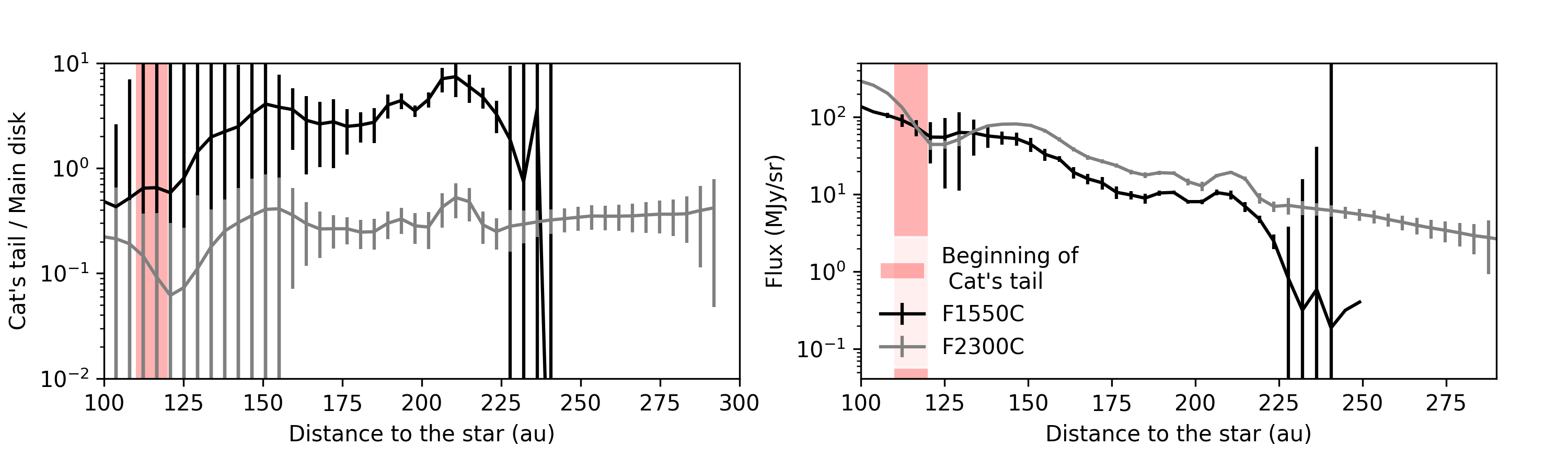}
    \caption{Cat's tail surface brightess in both MIRI filters. Left panel shows the contrast with the main disk, where the cat's tail at 15.5~\um (black line) is consistently brighter, sometimes closer to 2 orders of magnitude. Right panel shows the surface brightness of the cat's tail, higher at 23~\um (grey line), and decreasing constantly with the distance to the star. The red shaded area marks the beginning of the cat's tail feature.  }
    \label{fig:catstail_MD}
\end{figure*}

\subsection{Spectral energy distribution at different stellocentric distances}
\label{sec:subsec_scattered_SED}

We constructed SEDs, normalized in contrast relative to the star, combining the measurements from scattered light and thermal emission for different stellocentric distances, shown in Fig. \ref{fig:scattered_sed}. Starting from the disk contrast profiles obtained in Sect. \ref{sec:morphology}, the measurements for each individual distance were made averaging three pixels centered in each of the locations for each of the contrast profiles. For the full disk measurement, we integrated the full contrast profile. In scattered light, the disk brightness relative to the star decreases with wavelength from 2 to 4~\um (i.e. the disk-scattered light is blue relative to the star), reaching a minimum in contrast around 4--5~\um, and then rapidly brightening at longer wavelengths as thermal emission rises. 

The SED shows that the colors are consistent in both sides of the disk, and the spatial extent of the NE side is slightly larger, as we retrieve an additional data point at 250~au. 

\begin{figure}
    \centering
    \includegraphics[width=0.49\textwidth]{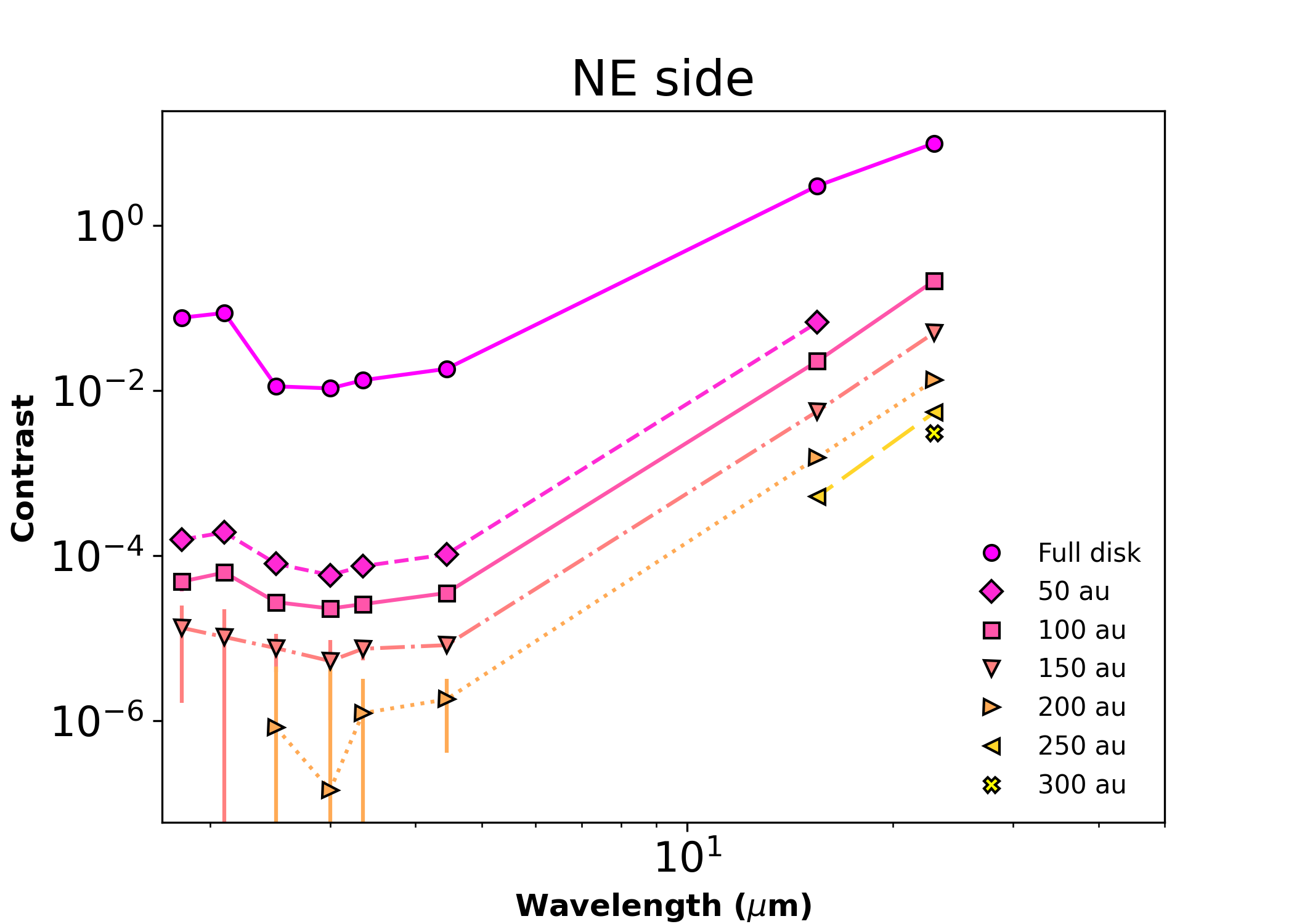}
    \includegraphics[width=0.49\textwidth]{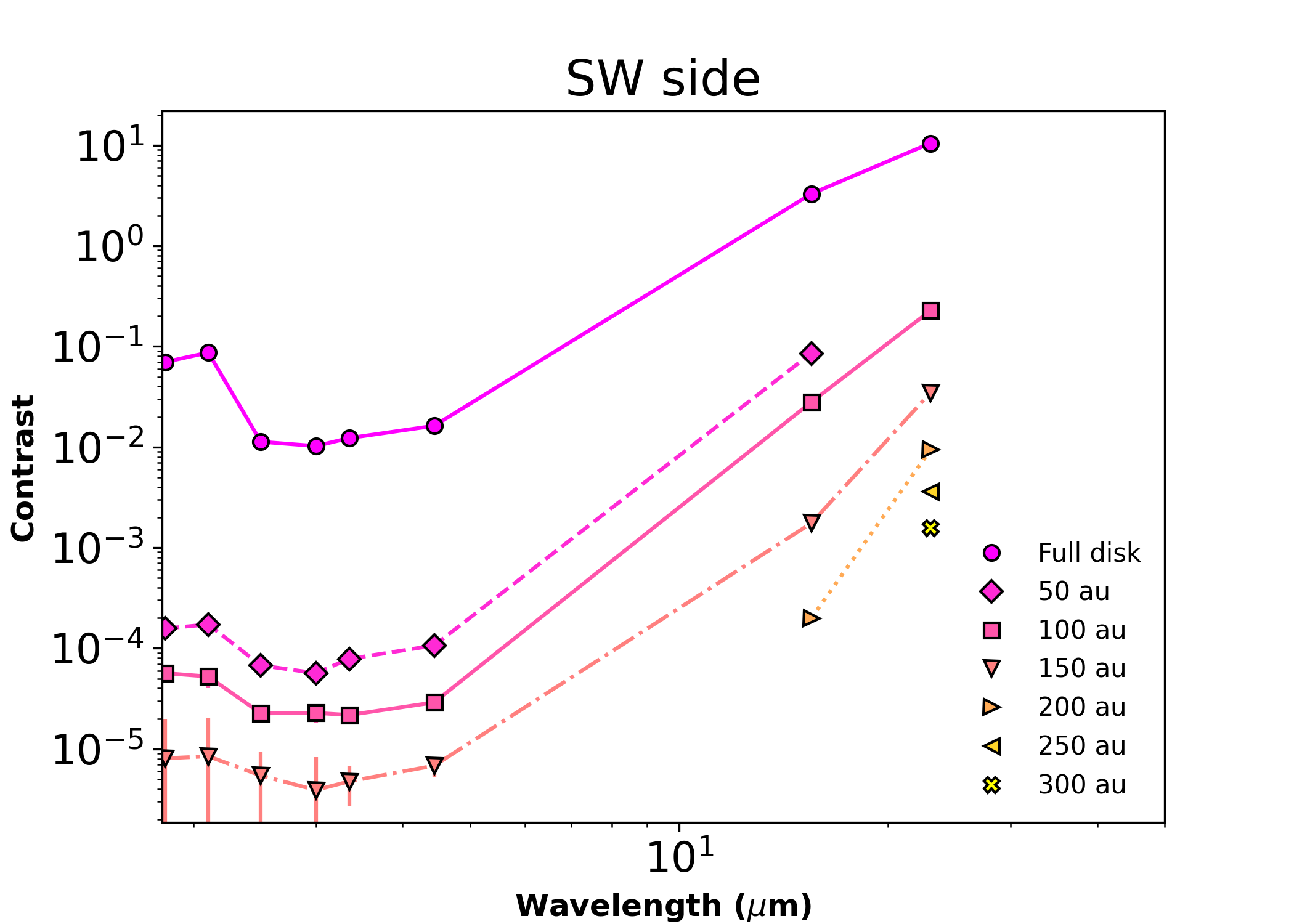}
    \caption{Contrast spectral energy distribution for thermal and scattered light in both sides of the disk. Data points have been obtained as explained in Sect. \ref{sec:subsec_scattered_SED}. Different colors and symbols refer to different positions on the disk, as labeled. Data below 2~\um, corresponding with NIRCam data, shows a decrease in the scattered light contrast; while 15.5~\um and 23~\um, corresponding with MIRI data, show how the thermal emission picks up as the brightness from the star decreases.}
    \label{fig:scattered_sed}
\end{figure}

\subsection{No evidence for water ice}
\label{sec:subsec_waterice}

Water ice, a fundamental ingredient for planet formation, is expected to sublimate in the inner regions of protoplanetary disks, and therefore be present only at long stellocentric distances in debris disks and mature planetary systems, past the so called snowline. Around a $\sim$1.8M$_{\sun}$ star, such as \bpic, the snowline is expected to be at $\sim$7~au (based on a simple black body temperature estimation for crystalline water ice). However, due to UV photo-dissociation, the evaporation of water ice can extend to 100-150~au \citep{Grigorieva07}.  %Given our IWA for NIRCam is $\sim$10~au, the spatial range of our observations is outside a hypothetical snowline.
%% Considering -75C the sublimaiton temperature for water ice. There is of course a dependance with composition, size of the grains, etc. but the goal here is to stablish that the snow line is inside the IWA, which is correct even for slightly different grains. ((1.8*const.R_sun/2)*(8052/198)**2.).to(u.au)

The set of NIRCam filters used was specifically chosen to provide coverage of the 3~$\mu$m water ice feature. This allows us to do an estimation of whether there is a significant amount of water at different  distances from the central star. Following Sect. 4.4 of \cite{Kim19}, we compared the surface brightness values obtained in and out of the expected feature range. Given we do not have observations at the wavelengths they selected, 2.8 and 3.2~\um, we chose filters at 2.5 and 3.35~\um, which are still expected outside and inside of a hypothetical water ice feature, respectively. Additionally, we repeated the calculation for 2.5 and 3.0\um, in order to verify our results. Simply calculating a ratio of surface brightness from Fig. \ref{fig:scattered_sed} at said wavelengths, we obtain values between 1.52$\pm$0.10 and 1.89$\pm$0.52 for distances between 50 and 150 au. Uncertainties were calculated from the uncertainty maps corresponding to the images obtained from the data reduction, and the corresponding error propagation.%Interestingly, the ratio drops to 0.77$\pm$0.03 at 200 au for the 2.5/3.35 ratio, and to 1.14$\pm$0.16 for the 2.5/3.0 ratio. 
Comparing to Fig. 16 in \cite{Kim19}, we see no evidence of water ice inside of 150 au. %, but not completely at 200 au. 
Given our method is purely based in photometric measurements, future IFU observations will provide a more precise determination (e.g. GO 1563, PI Chen).

%%%%%%%%%%%%%%%%%%%%%%%%%%%%%%%%%%%%%%%%%%%%%%%%%%%%%%%

\section{Sporadic dust production as a source for the disk's substructures}
\label{sec:dynamics_model}

The MIRI coronagraphic data reveal several features in the debris disk, described in detail in Sects. \ref{sec:morphology} \& \ref{sec:disk_color}: the ``fork," which appears from a radial extension of material on the NE side of the secondary disk; the ``cat's tail," which appears as a spatially extended, curved stream of material on the SW side of the disk; and nebulosity, especially to the west. These features exhibit unique traits, some not seen before in other disks:
\begin{itemize}
    \item The fork appears only to the NE side of the disk, suggesting an eccentric secondary disk.
    \item The cat's tail appears to curve away from the secondary disk, reaching an apparent position angle of $\sim45^{\circ}$ with respect to the main disk, then has a ``crook" near the tip that bends back toward the main disk.
    \item Over the extent of the cat's tail the surface brightness changes by a factor of $\sim5$ from base to tip, similar to the main disk (see Fig. \ref{fig:catstail_MD}).
    \item The nebulosity appears tendrilous, with apparent arcs of material curving off of the disk and tail.
    \item As shown in Figure \ref{fig:false_color}, all of these features appear blue compared to the main disk, suggesting a common origin that differs in composition or grain size distribution and is hotter than the main disk. 
\end{itemize}
These unique features motivate the need for a dynamical explanation and can help constrain models of different dynamical scenarios. Here we examine several possible causes. Motivated by the extended secondary disk that is seen only on the NE side of the disk, we start by discussing a steady-state collision point model that naturally produces eccentric disks, but ultimately does not adequately explain the color, flux, or crook of the cat's tail. We then perturb this model by invoking sporadic dust production and show that it can reproduce all of the features described above.

\subsection{Steady state collision point model}
\label{sec:steady_state_model}

\begin{figure*}
    \centering
    \includegraphics[width=\textwidth]{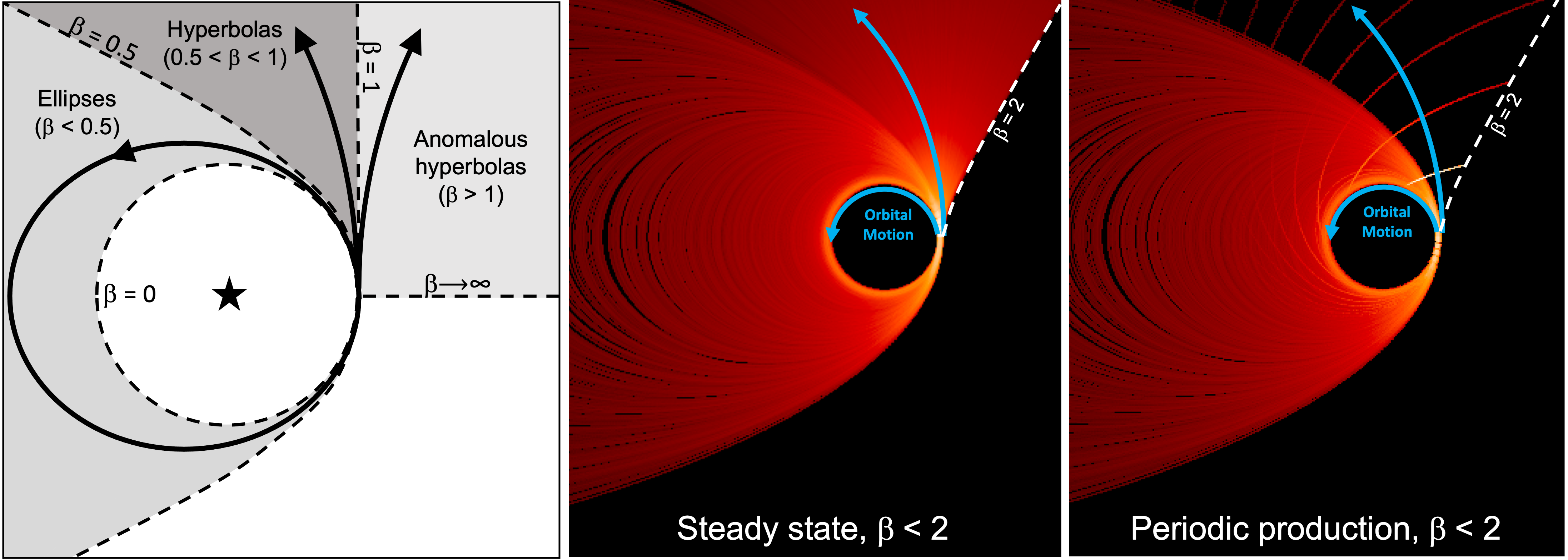}
    \caption{Left: orbital trajectories of grains from a collision point as a function of $\beta$ (inspired from Figure 1 of \cite{Krivov10}). Middle: steady-state collision point model with $\beta<2$ producing an eccentric disk of dust and a fan of blow-out grains. Right: collision point model with periodic dust production events at 100 year intervals. Individual dust production events leads to curved streams of material that travel outward while forming a spiral, with bound grains eventually populating the main disk.}
    \label{fig:dust_dynamics}
\end{figure*}

The appearance of an asymmetric extension on only one side of the observed disk suggests the presence of an eccentric secondary disk. Indeed, Figure \ref{fig:false_color} shows a population of bluer (hotter) grains that appear to reside in this eccentric disk. Generating an eccentric disk of dust is not dynamically challenging---in fact, at some level, it's unavoidable. The collision point models of \cite{Jackson2014} predict that a single massive collision will, over time, lead to a belt of material that continually returns to the same ``collision point," where small dust grains are continuously produced. The dynamics of these small dust grains are controlled by $\beta$, the ratio of radiation pressure force to the gravitational force. The left panel in Figure \ref{fig:dust_dynamics}, inspired by Figure 1 from \cite{Krivov10}, illustrates the trajectory of grains as a function of $\beta$ when launched from a progenitor on a circular orbit. Bound dust with $\beta<0.5$ is immediately blown onto orbits with larger semi-major axes and eccentricities, resulting in an eccentric, asymmetric halo of small dust with periastron located at the collision point (see lower panel of Figure 15 in \citealp{Jackson2014}). These collision point models have been recently invoked to explain the morphology of a multitude of debris disks seen in scattered light~\citep{Jones23}. We note that over long timescales, collisions will eventually resorb the eccentric disk \citep{Kral2015}. Its presence should therefore place an upper limit on the age of the original destructive event, which we do not model in this paper.

As shown in Figure \ref{fig:dust_dynamics}, unbound grains with $\beta>0.5$ leave the system following hyperbolic or anomalous hyperbolic orbits on relatively short timescales. When continually produced, these unbound grains produce a fan of material leaving the disk (see upper panel of Figure 15 in \citealp{Jackson2014}). The middle panel of Figure \ref{fig:dust_dynamics} shows an example of steady state dust production from a collision point model. For this illustrative example, we assumed a maximum value of $\beta=2$ (ultimately this is determined by the properties of the star and dust grains) and assumed all dust production occurred at the same point, originating from parent bodies on circular orbits.

For this steady-state scenario, we posit that the extended secondary disk to the NE, the HST-observed inner warp/secondary disk \citep{Heap2000,Golimowski06}, and the CO clump \citep{Dent2014} all belong to a single secondary disk as suggested by \cite{Apai15}. The CO clump may trace the location of the collision point, as suggested by \cite{Jackson2014}, which would place periastron to the SW side of the disk and the eccentric halo to the NE, producing the observed extension of the secondary disk on that side.  

\begin{figure}
    \centering
    \includegraphics[width=\columnwidth]{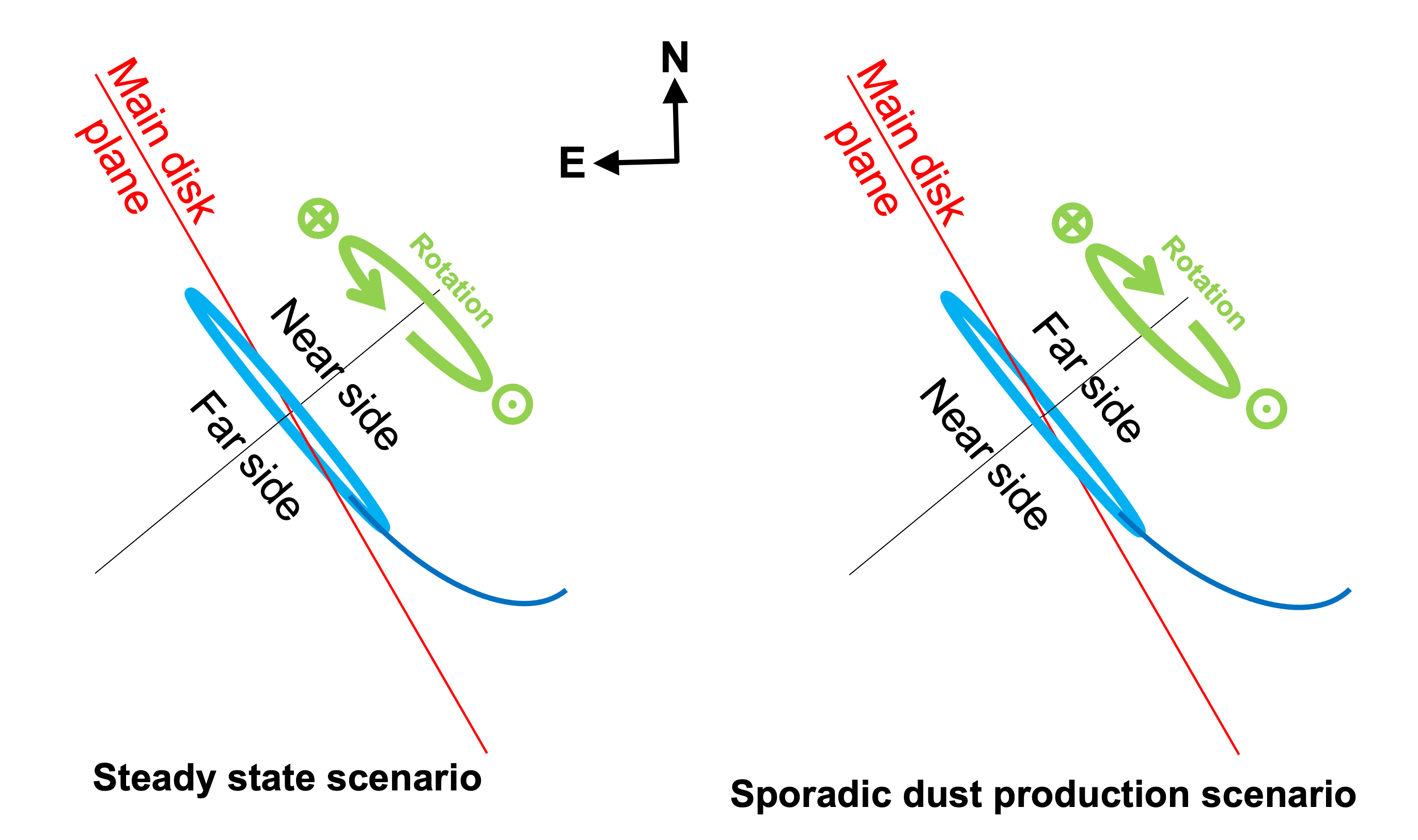}
    \caption{Orientation of the secondary disk for the steady-state model and our favored sporadic dust production model, which each require a slightly different inclination of the secondary disk. Note that the difference in assumed orientation is for the \textit{secondary disk only}; in both cases the main disk orientation remains nearly edge-on with the NW side nearer to us, as has been inferred from brightness asymmetry attributed to forward scattering \citep[e.g.][]{Golimowski06, Milli2014, Millar-Blanchaer15}. The rotation of both disks (main and secondary) is such that the SW side's orbital motion is toward us, unambiguously known from radial velocity observations of the CO gas \citep{Dent2014} and the planet \BPic b \citep{Snellen2014Natur.509...63S}. }
    \label{fig:orientation_illustration}
\end{figure}

Bound grains can only explain the appearance of the extended secondary disk. To explain the cat's tail and nebulosity, we must invoke unbound grains. In this scenario, blowout grains emanate from the collision point on hyperbolic orbits that curve toward the observer. If the secondary disk were seen edge-on, these orbits would simply appear radial, as all forces invoked so far are radial. However, if we invoke a small inclination ($\sim6^{\circ}$) for the secondary disk relative to the main disk and ensure a component of the line of nodes is within the sky plane, such that the secondary disk is not edge-on from our perspective, then the curvature of the hyperbolic orbits can create the observed apparent position angle of $\sim45^{\circ}$. The nebulosity and cat's tail could therefore be explained as belonging to a fan of blowout grains. Importantly, this model suggests that while the base of the cat's tail is at $\sim120$ au from the star, the tip of the cat's tail is at a physical distance of $\sim1000$ au. The left panel of Figure \ref{fig:orientation_illustration} illustrates the orientation and rotation of the disk and cat's tail in this steady-state collision point scenario.

\begin{figure*}[t]
    \centering
    \includegraphics[width=\textwidth]{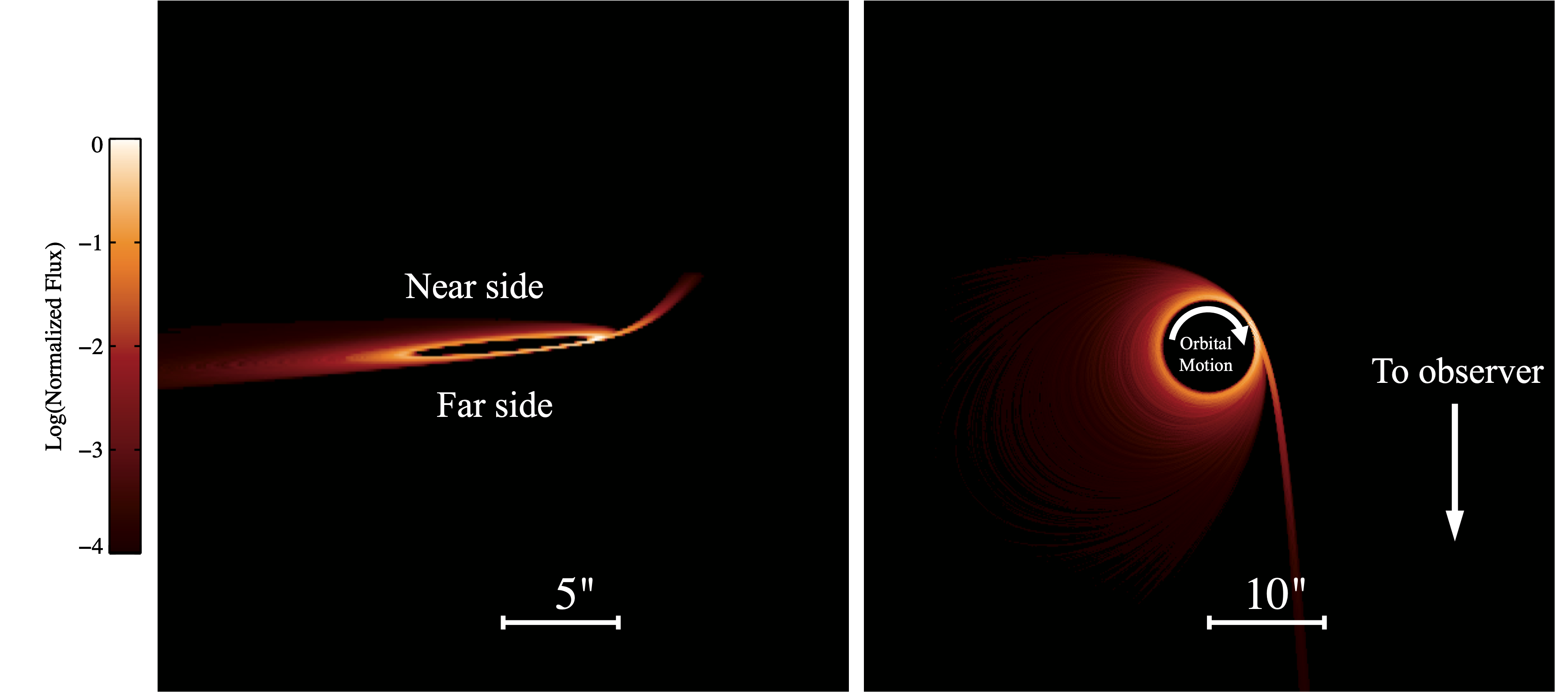}
    \caption{Synthetic 23 $\mu$m thermal emission image of a two-component ``toy" steady-state collision point model with $\beta=0.7$ blowout grains and without external forcing (e.g., ISM drag), as discussed in section \ref{sec:steady_state_model}. The left panel shows the projected view oriented the same as Figure \ref{fig:nebulosity_annotated}, and the right panel shows a face-on view looking down onto the disk from the NW. Invoking preferential production of $\beta\sim0.7$ via porous silicates can produce a cat's tail-like stream of material, but seems to be at odds with the surface brightness profile, color, and crook of the cat's tail.}
    \label{fig:steady_state_model}
\end{figure*}

Figure \ref{fig:steady_state_model} shows a synthetic thermal emission image of a ``toy" model of this scenario, as applied to \BPic. We assumed parent bodies on circular orbits occupying a ring with 10\% fractional width. At a single azimuthal location in the parent body ring, we launched bound grains with $\beta$ uniformly and randomly distributed from $0.01$ to $0.49$. We integrated their orbits using a leap-frog integrator for a full orbital period, periodically recording their locations to simulate constant production. We assumed $\beta \propto s^{-1}$, where $s$ is grain radius, and weighted them according to a Dohnanyi size distribution \citep{Dohnanyi1969}. We then used {\tt dustmap} \citep{stark2011} to synthesize the 23 micron thermal emission image shown in Figure \ref{fig:steady_state_model}, adopting optical constants predicted by Mie theory for 95\% porosity astronomical silicates \citep{LiGreenberg97}. For the cat's tail feature, we adopted a stream of grains with $\beta\sim0.7$. The disk is inclined by 4$^{\circ}$ from edge-on with the line of nodes rotated out of the sky plane by 30$^{\circ}$ and has a position angle of 5$^{\circ}$. The right panel of Figure \ref{fig:steady_state_model} shows the same model face-on. We note that we arbitrarily scaled the number of $\beta\sim0.7$ grains relative to the bound dust. In reality, this is not allowed, as the relative surface brightness is governed by the size distribution of collisional fragments and collision rate of the disk. Prior modeling efforts suggest a steep drop in the size distribution at the blow-out size \citep{ThebaultKral2019}.

We identify four challenges for this steady state model. First, to explain the cat's tail we must invoke a near-singular value of $\beta$ for all blowout grains. Adopting a range of $\beta>0.5$ would result in a broad fan-like feature that does not resemble the observed stream. One potential explanation for this is that a broad range of grain sizes may \emph{all have the same value} of $\beta\sim0.7$. \cite{Arnold2019} determined the $\beta$ value for different grain sizes and compositions of dust around a \BPic like star using both Mie theory and the discrete dipole approximation method.  Figure 4 in \cite{Arnold2019} shows that Mie theory predicts highly-porous silicates have a near-constant $\beta\sim0.7$ over several orders of magnitude in grain size. While in theory this should create a stream of material resembling the cat's tail, in practice we find this unlikely, as $\beta$ is inversely proportional to density, and thus small variations in density among grains would diffuse this stream substantially.

Second, the stretch of our synthetic image shown in Figure \ref{fig:steady_state_model} is substantially larger than the data set shown in Figure \ref{fig:nebulosity_annotated}; the base of the cat's tail is much brighter than the tip in our model. Put simply, porous silicate grains at 1000 au are substantially cooler and fainter in thermal emission than at 100 au. Future studies may find a mixture of materials, which we do not investigate here, that can mitigate this issue.

The third challenge of our model is the crook near the tip of the cat's tail. While our model explains the apparent curvature \emph{away} from the main disk, it does not explain the observed curvature back \emph{toward} the main disk. To address this, we must invoke either a difference in initial conditions, such as a mass dependent initial velocity distribution for collisional ejecta, or some sort of external forcing, such as ISM drag~\citep{Debes09}. Indeed, the proper motion of \Bpic is primarily toward the north\footnote{From Hipparcos: pm$_\mathrm{RA}= 4.65$ mas/yr; pm$_\mathrm{Dec}= 83.10$ mas/yr}, potentially providing an ISM drag force that points roughly parallel to the main disk midplane. Further, the highly-porous silicates that could explain the selective $\beta$ production may have low enough density to be altered by the ISM. However, the porous silicates in the cat's tail would occupy a broad range of grain sizes, such that ISM drag would splay the tip of the cat's tail, effectively size-sorting it into a fan-like structure, which is not observed.

Finally, the cat's tail appears to have a finite extent, with its flux dropping rapidly at the tip. In a steady-state scenario, the cat's tail should extend to much greater distances and the flux of the cat's tail should drop smoothly until it falls below the sensitivity limits of our observations.

\subsection{Sporadic dust production model}
\label{sec:sporadic_model}

Here we present our preferred model that addresses all of the above shortcomings of the previous model by invoking a simple change: sporadic dust production, in which dust is produced at irregular intervals from irregular locations within the disk. We start by discussing the concept of discretized dust production events.

When produced continuously, bound dust forms a broad eccentric disk and blowout grains form a fan, as shown in the middle panel of Figure \ref{fig:dust_dynamics}. However, we should expect that, at some level, a disk is generated by many individual collisional or outgassing events. The right panel of Figure \ref{fig:dust_dynamics} shows the expected geometry of a collision point model in which dust is produced periodically in regular, discretized intervals. Here the geometry of an individual production event appears as an arc of debris spreading outward. Importantly, this arc is $\beta$-sorted, such that particles near the base have $\beta\sim0$ while particles near the tip are unbound.

To explain the cat's tail and tendrilous nebulosity observed in the MIRI data set, we posit a quasi-steady-state eccentric secondary disk with \emph{sporadic} dust production events. In our model a single recent collision does not produce the entire secondary disk. Rather, the secondary disk (including its extension to the ``fork") may be relatively old, the product of many evolved dust production events over its history as predicted by the collision point model, while the cat's tail, nebulosity and other asymmetric features like the CO clump are due to recent individual dust production events. While dust production must occur to the SW on average to explain the eccentricity of the secondary disk, we do not require a single point of production when modeling the cat's tail or nebulosity. Rather, we allow dust to be produced sporadically in time and location. By doing so, we are able to reproduce the geometry and finite extent of the cat's tail, as well as arcs that roughly resemble some of the observed nebulous features.

From the perspective of an observer, the geometry of a sporadic dust arc could take on many forms, from linear features to eccentric spirals, depending on the distribution of $\beta$ values, orbit/orientation of the progenitor, and time since the collision. To investigate these structures, we integrated the equations of motion for 10k particles covering a broad range of $\beta$ values, releasing them from different orbits and orientations and tracking their evolution with time. We examined inclinations for the secondary disk ranging from $0-10^{\circ}$ and position angles ranging from $0-6^{\circ}$ with respect to the main disk. Given that the longitude of ascending node cannot point toward the observer, as this would produce an edge-on secondary disk with strictly radial features, we limited the line of nodes to lie in the sky plane---future modeling efforts could explore this additional free parameter.  In our efforts to reproduce a feature resembling the cat's tail, we found the following:
\begin{itemize}
    \item Assuming the secondary disk orbits in the same direction as the CO gas \citep{Dent2014}, we must swap the near and far sides of the secondary disk, as shown in the right panel of Figure \ref{fig:orientation_illustration}. This change affects the secondary disk only, not the main disk, and is a relatively minor change in inclination given the nearly edge-on alignment of the disks.
    \item Assuming the line of nodes of the secondary disk lies in the sky plane, the original point of dust production must occur on the far side of the disk near the secondary disk's projected minor axis; the orbital anomaly of the collision point is likely within $\sim5$ degrees of the projected minor axis.
    \item We require a maximum $\beta$ of $\sim10$ to reproduce the curvature and extent of the cat's tail. These high-$\beta$ grains move outward faster, which allows grains to reach $\sim1000$ au in a fraction of the progenitor's orbital period, producing the proper curvature and extent of the cat's tail. 
\end{itemize} 

The left panel of Figure \ref{fig:SEDs} shows $\beta$ calculated as a function of grain size for several commonly-invoked compositions and porosities using Mie theory and the Bruggeman mixing rule. Dotted lines show non-porous dust while solid lines show dust with 95\% porosity. The dotted red curve shows the best fit model to the main disk's halo component from \cite{Ballering16}, while blue, orange, and black lines show organic refractory~\citep{LiGreenberg97}, astronomical silicate~\citep{DraineLee1984}, and amorphous carbon grains~\citep{Zubko1996}, respectively. To first order, high porosity shifts these curves to larger grain sizes and broadens the peak, resulting in larger and more abundant blowout grains; the fact that these grains are visible at 23 $\mu$m suggests that they may have high porosity. To achieve $\beta\sim10$ dust for \BPic, we likely must invoke refractory organics or amorphous carbon. We emphasize that independent of color and flux, \emph{the shape of the cat's tail} appears to inform the dust composition.

\begin{figure*}
    \centering
    \includegraphics[width=\textwidth]{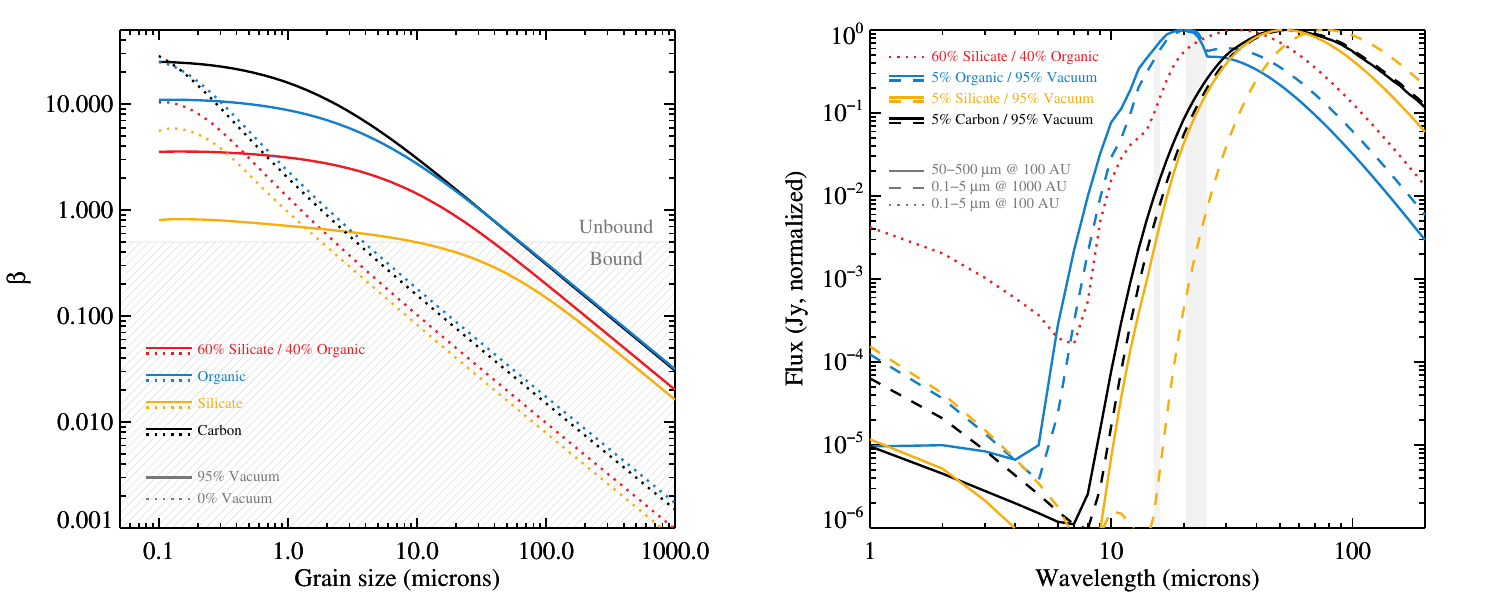}
    \caption{Left: $\beta$ as a function of grain size for \cite{Ballering16} best fit composition (dotted red) and other compositions/porosities. Our dynamic model requires high $\beta$ and larger grains, consistent with fluffy refractory organic material. Right: Normalized spectral energy distributions for the best-fit \cite{Ballering16} halo component for the main disk (red), as well as three highly porous compositions. Only fluffy organic refractory material can produce a bluer mid-IR color (higher ratio of F1550C/F2300C) than the \cite{Ballering16} composition. Solid, dashed, and dotted lines indicate different particle size ranges and/or locations, as labeled in gray.}
    \label{fig:SEDs}
\end{figure*}

In addition to requiring a high $\beta$ value, the dust must appear bluer than the main disk. The right panel of Figure \ref{fig:SEDs} shows the spectral energy distributions for highly porous (i.e., ``fluffy") grains around \bpic calculated via Mie theory. Solid lines correspond to large grains (50--500 $\mu$m) near the base of the cat's tail (100 au), while dashed lines correspond to small grains (0.1--5 $\mu$m) near the tip of the cat's tail (1000 au). The dotted red line shows the best fit composition to the main disk halo from \cite{Ballering16}, valid only for small grains at 100 au. Vertical gray bars mark the approximate bandpasses of the F1550C and F2300C filters. Only one of the compositions (blue curves) allows the dust to appear bluer than the main disk (red curve): organic refractory material. We therefore focus solely on fluffy organic refractory material for our sporadic dust production model.

Prior to modeling multiple sporadic events, we first focus on reproducing the single most prominent feature: the cat's tail. 
Figure \ref{fig:sporadic_no_ISM} shows a synthentic 23 $\mu$m thermal emission image of a two-component sporadic dust production model without any external forcing. To produce the eccentric disk component, we simulated bound grains originating from a single progenitor on a circular orbit at 85 au. 
We adopted $0.01<\beta<0.49$ and integrated the equations of motion for an orbital period, calculated their grain size from the curve shown in Figure \ref{fig:SEDs}, and weighted them according to a Dohnanyi size distribution at the collision point. 
We repeated this process for the cat's tail component, with $0.5<\beta\lesssim10$, and adopted a steeper-than-Dohnanyi size distribution $dN/ds\propto s^{-3.8}$ for the ejecta. This size distribution better reproduces the flux over the extent of the cat's tail and is consistent with the aftermath of recent dust production event that has not yet reached collisional equilibrium \citep{leinhardtstewart2012}. For the cat's tail, we launched grains from an orbital anomaly of $5^{\circ}$ to the NE of the secondary disk's apparent minor axis and stopped the orbital integration after 150 years. We adopted an inclination of $6^{\circ}$ from edge-on and a position angle of $2.5^{\circ}$ for both components. 

\begin{figure*}
    \centering
    \includegraphics[width=\textwidth]{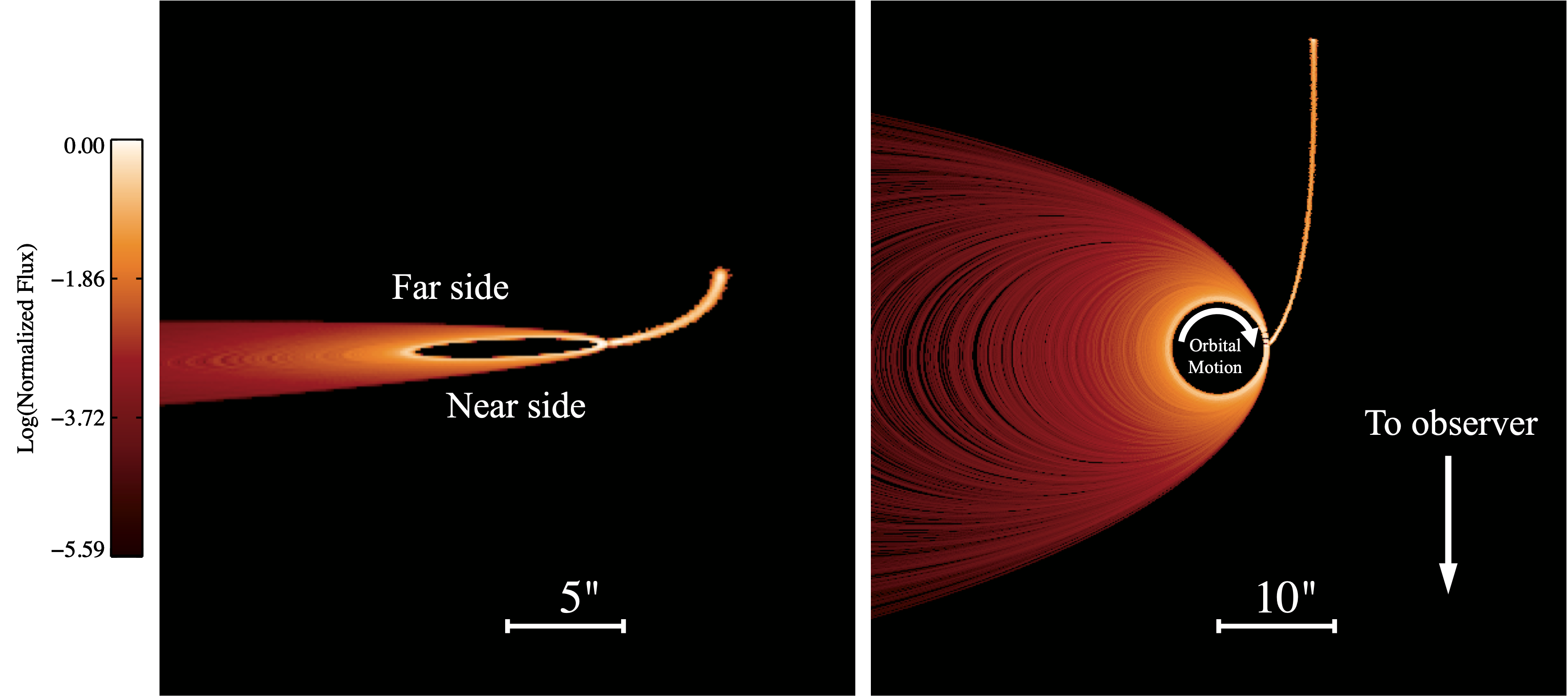}
    \caption{Synthetic 23 $\mu$m thermal emission image of a two-component model without external forcing, as proposed in section \ref{sec:sporadic_model}, consistent of a quasi-steady-state secondary disk and a single 150 yr old collisional event. This reproduces the shape of the cat's tail, with exception to the crook near the tip. The left panel shows the projected view oriented the same as Figure \ref{fig:nebulosity_annotated}, and the right panel shows a face-on view looking down onto the disk from the NW.}
    \label{fig:sporadic_no_ISM}
\end{figure*}

As shown in Figure \ref{fig:nebulosity_annotated}, the cat's tail features a crook near the tip, in which it bends back toward the main disk plane. To explain this crook, we must include additional physics in our model. We see two possible options. First, we could include external forcing primarily directed toward the SW, such as an ISM drag component~\citep{Debes09}. Alternatively, we could posit that the collisional or outgassing event produced a size-dependent velocity distribution in the SW direction, such that small grains near the tip of the cat's tail received a much larger velocity kick than their larger counterparts. Size-dependent ejection velocity is a known outcome of cometary outgassing, as small grains are quickly entrained by the gas tail and accelerated at a higher rate until the gas tail dissipates \citep[e.g.,][]{Moorhead2021}. Here we focus on the former, ISM drag, as an example, though we do not rule out alternative explanations.

Figure \ref{fig:sporadic_with_ISM} shows a synthetic 23 $\mu$m thermal emission image for our preferred sporadic dust production model including ISM drag. We adopted the same parameters as our previous model, but added ISM drag, which extends the tip of the cat's tail to the SW. Here we have included multiple sporadic dust production events that qualitatively reproduce some of the geometry of the observed nebulosity and arc-like features seen in the MIRI data, as marked in Figure \ref{fig:nebulosity_annotated}. These fainter ``tendrils" cover a broad range of times since collision and are all generally created near the projected minor axis, which notably corresponds to roughly where the primary and secondary disk planes cross. Events consistent with the SW nebulosity occur on the far side of the disk and are $\sim150-200$ years old, while events consistent with the arc-like features to the NE occur on the near side of the disk and are $\gtrsim1000$ years old. We note that all ages are approximate, as they are somewhat degenerate with the orbital period of the parent body and $\beta$ values for the dust.  

The ISM drag we modeled can successfully replicate the crook in the cat's tail. Unlike the steady-state model shown in Figure \ref{fig:steady_state_model}, ISM drag only affects the tip of the cat's tail because the stream is size-sorted with smaller grains near the tip. For the model shown in Figure \ref{fig:sporadic_with_ISM}, we adopted a density of $0.09$ g cm$^{-3}$ for our fluffy refractory organic grains (5\% of 1.8 g cm$^{-3}$; see \cite{ZubkoDwekArendt2004}). Such a low grain density leads to potentially plausible ISM parameters. For Figure \ref{fig:sporadic_with_ISM}, we adopted an ISM density of 30 H atoms cm$^{-3}$ and a relative ISM velocity of 4 au/yr in the $+x$ direction and 1 au/yr in the $+z$ direction, where $+x$ points to the SW along the secondary disk's midplane, and $+z$ points to the NW along the normal to the secondary disk's midplane. This ISM velocity is $\sim-3$ au/yr both in the directions of right ascension and declination, broadly consistent with the proper motion of \bpic. We did not attempt to exhaustively explore the ISM drag parameter space and simply note that ISM density and relative speed are somewhat degenerate; ISM density can be decreased for a relatively small increase in relative speed. We also note that we did not include circumstellar gas drag, which may complicate our model.

\begin{figure*}
    \centering
    \includegraphics[width=\textwidth]{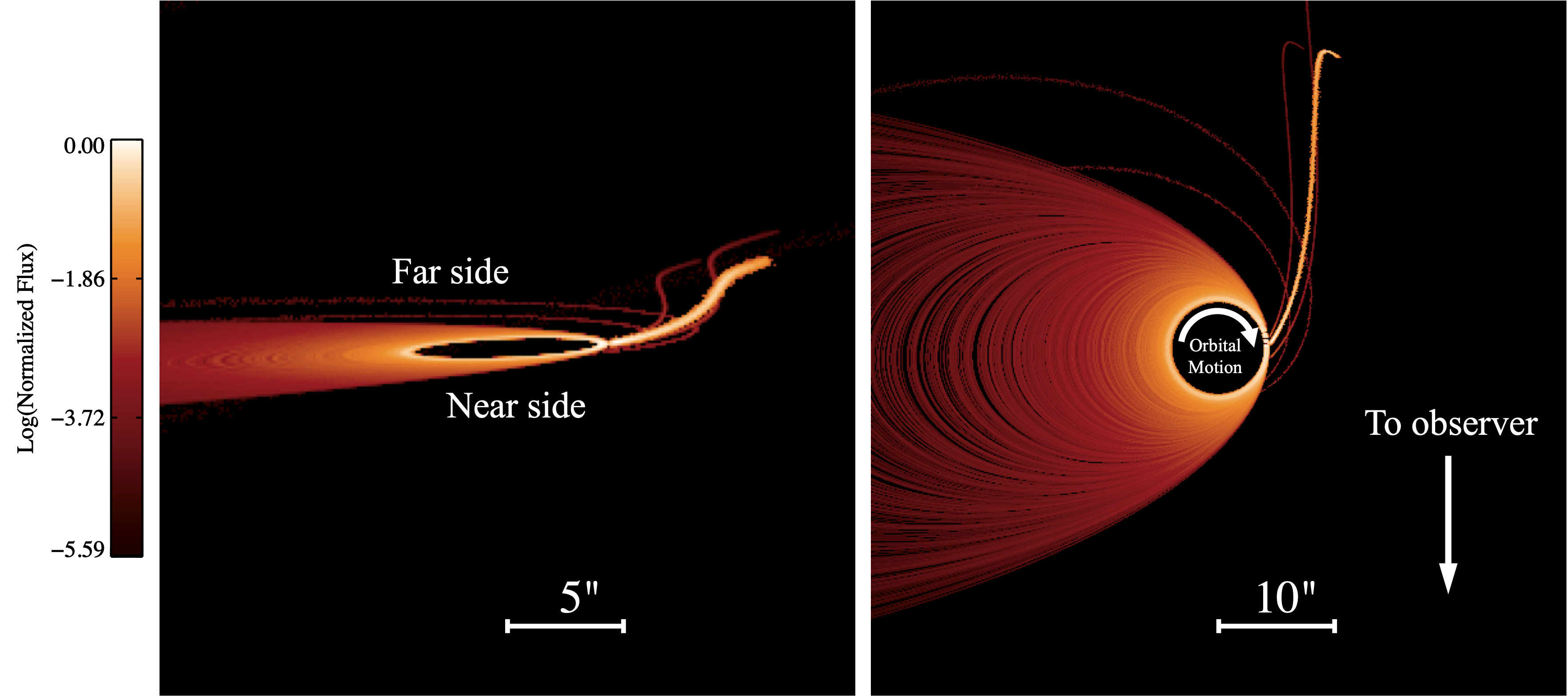}
    \caption{Synthetic 23 $\mu$m thermal emission image of a multi-component sporadic dust production model with an ISM drag force, which broadly reproduces the shape, color, and flux of the cat's tail ($\sim150$ years old), as well as the presence of additional arcs to the SW ($\sim150-200$ years old) and NE ($\gtrsim$1000 years old), as marked in Figure \ref{fig:nebulosity_annotated}. The left panel shows the projected view oriented the same as Figure \ref{fig:nebulosity_annotated}, and the right panel shows a face-on view looking down onto the disk from the NW.}
    \label{fig:sporadic_with_ISM}
\end{figure*}

The same porous refractory organic grains invoked to explain the shape of the cat's tail, as well as the crook in the cat's tail, also help to explain the flux and color of the cat's tail. To measure the flux along the cat's tail, we had to isolate and subtract the ``background" disk flux (i.e., the main disk) from the cat's tail. To do so, we created a mask that followed the peak of the cat's tail with a half width of 3 pixels. We estimated backgrounds on a per-pixel basis within this region by interpolation using a high-order two dimensional polynomial fit to the regions located 6-12 pixels on either side of the cat's tail. After subtracting backgrounds, we measured the F1550C and F2300C average surface brightness along the cat's tail in steps of $0.44\arcsec$. We find that the F1550C and F2300C surface brightness decreases by a factor of $4-5$ from a projected separation of $8.25\arcsec$ to $11.4\arcsec$. 

We then used {\tt dustmap} to synthesize F1550C and F2300C images of our cat's tail model shown in Figure \ref{fig:sporadic_with_ISM}. We used Mie theory and adopted optical constants appropriate for 95\% porous organic refractory material~\citep{LiGreenberg97}. We approximated the bandpass-integrated flux of each filter by integrating over 100 individual wavelengths. We approximated each bandpass as a top-hat function ranging from $15.1-15.9$ $\mu$m and $20.5-25.0$ $\mu$m for the F1550C and F2300C filters, respectively. No background subtraction was needed for the model, as the cat's tail component was modeled in isolation. 

The left panel of Figure \ref{fig:cats_tail_color} shows the F2300C flux as a function of separation for the main disk midplane as well as the background-subtracted cat's tail. Empty blue circles show the cat's tail without deconvolution, while filled blue circles show the cat's tail with deconvolution. The black line indicates the model flux. While our model does not reproduce the shape of the cat's tail flux as a function of separation, it is on par with the total decrease in flux from base to tip.

The right panel of Figure \ref{fig:cats_tail_color} shows the ratio of F1550C to F2300C surface brightness along the cat's tail for the convolved and deconvolved reduced data sets, with the main disk's color shown for reference. The black line indicates the color of our sporadic dust production cat's tail model. Porous organic refractory grains produce a color on par with, or in excess of the observations over the entirety of the cat's tail. This is due to the excess heating of small grains, as evidenced by the solid and dashed blue lines in Figure \ref{fig:SEDs}. The differences between the model and data in Figure \ref{fig:cats_tail_color} suggest that further modeling efforts are needed to better constrain composition, porosity, and grain size distribution, and that the composition may be a layered mixture of organic refractory material and silicates as suggested by \cite{LiGreenberg98}.

\begin{figure*}
    \centering
    \includegraphics[width=\textwidth]{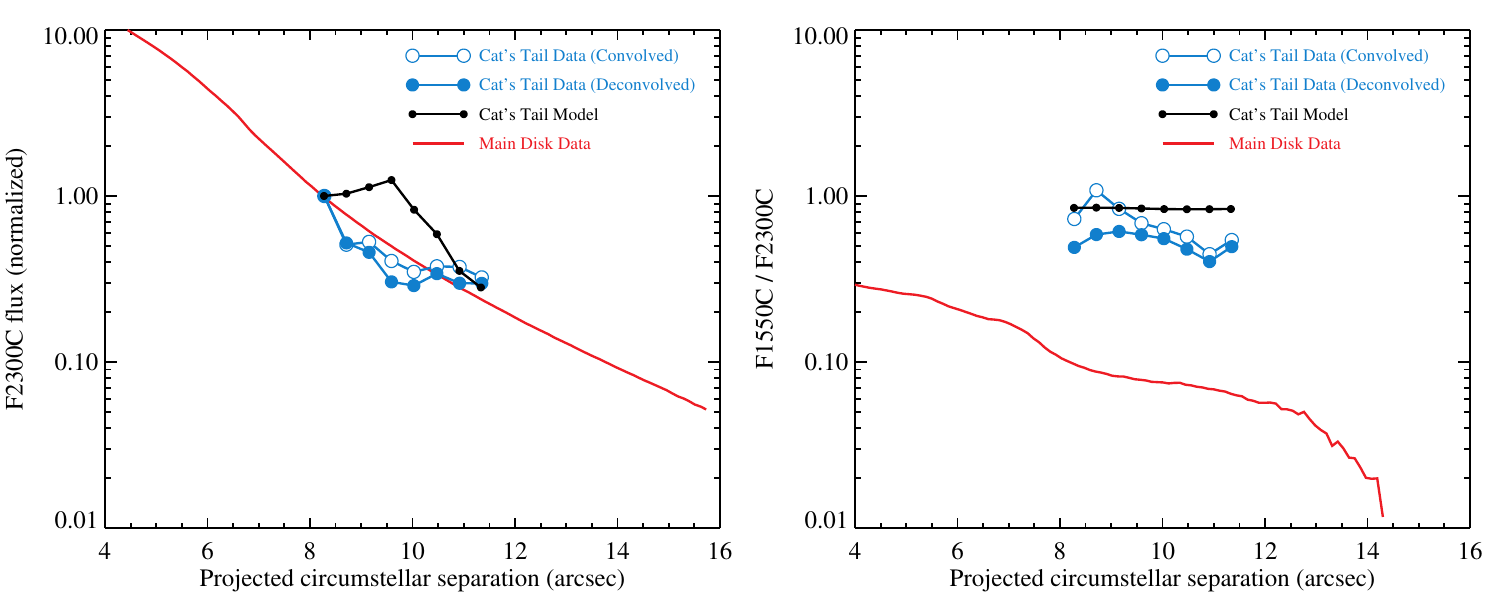}
    \caption{Left: F2300C flux as a function of apparent separation measured along the main disk's midplane, as well as the measured and modeled cat's tail. Right: Color (F1550C/F2300 flux ratio) of the components. Our model with porous organic refractory grains largely reproduces the flux and color over the length of the cat's tail.}
    \label{fig:cats_tail_color}
\end{figure*}

To estimate the mass of the cat's tail, we scaled our optically-thin model such that the total flux of the cat's tail matched observations. We adopted a density of $0.09$ g cm$^{-3}$, appropriate for  highly porous refractory organic material, and calculated the size-integrated mass. We find that our model requires $2\times10^{19}$ kg of dust smaller than 1 mm, or roughly the equivalent of a large main belt asteroid in the solar system. A total mass estimate for the original parent body is highly uncertain, as we have no constraints on the size distribution or density of larger objects. However, extrapolating that size distribution while maintaining the density for porous grains, and assuming only 10\% of the original object is fragmented, suggests a total mass potentially $\sim100$ times larger, corresponding to the mass of a small moon in the solar system. The fainter tendrils to the SW and NE have $\sim5-100$ times less flux than the cat's tail and would require correspondingly less mass.

\subsection{Collision with a body on a highly inclined orbit}

We note that one seemingly obvious explanation for the apparent high inclination of the cat's tail could be simply that the progenitor was on a highly inclined orbit. We cannot rule out this scenario. However, in addition to invoking an unseen object on a highly inclined orbit, this scenario has several shortcomings. We would expect a collision of this nature to travel outward as a detached cloud. To remain attached to the main disk over $\sim100$ au, the collision would have to produce a cloud of material with a very broad velocity dispersion. Further, the observed CO emission appears to reside within the secondary disk. If created by a recent collision with a highly inclined progenitor, we would expect the CO clump to show a more significant out-of-plane extension. Finally, it is not clear how this scenario could explain the complex tendrilous nature of the surrounding nebulosity. 

\section{Discussion}
\label{sec:discussion}

\subsection{Structure and morphology}

Forty years of study have yielded an extensive literature about the morphology and composition of the \bpic system. 
Observations in scattered light, both from ground \citep[e.g.][]{Golimowski93,Larwood01,Milli2014,Esposito20} and space \citep[e.g.][]{Burrows95,Heap2000,Golimowski06,Apai15} have characterized the disk. Almost all of them recover the presence of a warp, linked to an inclined secondary disk component, and an asymmetry in the surface brightness profiles of the NE and SW sides of the disk. Within the inner few arcseconds the NE side of the disk is fainter, but at larger separations the NE side is brighter and more extended (a result which can be traced all the way back to the original pioneering observations of \citealp{Smith84}, and is seen especially clearly in the deep VLT/FORS1 images of \citealp{Janson21}). The scattered light observations with NIRCam are broadly consistent with these prior results. 

There are some structures, like the SW dust and CO/C gaseous clumps at around 50-60~au, that are only visible in thermal emission, and are therefore only recovered at longer wavelengths, i.e. mid-IR \citep{Telesco05} and sub-millimeter \citep{Dent2014,Cataldi18}. 
In recent work, \citet{Skaf23} identified several ``clumps" in mid-IR images at locations closer than 5\arcsec~to the star. They suggested the main clump, which they locate at $\sim$52-56~au, has moved when comparing observations spanning 16 years. However, another paper by \cite{Han23} suggest that the change in position is not significant considering the uncertainties. Our results (see Fig. \ref{fig:asymmetry_JWST}) indicate the asymmetry between the NE and SW sides of the disk has a clumpy structure, but however the location of the clumps we detect is not coincident with the ones reported in \cite{Skaf23} with the exception of their C1 clump. This corresponds to the brightest clump we observe in both MIRI filters. The measurements of peak position described in Sect. \ref{sec:subsec_surfacebrightness} indicate that the location is wavelength dependant, with the clump at 23~\um being located further from the star. This would be consistent with each wavelength tracing material at different temperatures, althoug we should not rule out the possibility of instrumental artifacts, given the location of the F2300C mask. Our measurement at 15.5~\um is fairly consistent with the locations reported in \cite{Skaf23} and \cite{Han23}. Fig. \ref{fig:asymmetry_JWST} includes shaded regions with some of the estimations from \cite{Skaf23} and \cite{Han23}. 

In the same region, millimeter observations with ALMA by
\cite{Dent2014} and later \cite{Cataldi18} revealed the presence of CO and \ion{C}{1} gas peaks, respectively, as well as a significant asymmetry in the continuum, potentially corresponding to the hypothesized dust clump. Analysis of the gaseous component in \cite{Dent2014} revealed that the gaseous disk is inclined with respect to the continuum counterpart by $\sim$5$^\circ$, similar to the findings for the secondary disk \citep[e.g.][]{Heap2000,Golimowski06}. Their velocity analysis also showed the rotation of the disk, with the SW side approaching the observer, and the NE side moving away; as shown in Fig. \ref{fig:orientation_illustration}. More recently, the re-analysis of ALMA band 6 data \citep{Matra19}, showed no evidence of a NE/SW asymmetry. While they claim that their result is consistent with no evidence of large dust grains causing the asymmetry due to the lack of detection at 24~\um in \cite{Telesco05} and in \cite{Vandenbussche10}, there is a possibility that the PSF FWHM in those works is too large to characterize the dust clump. Further JWST MIRI observations could investigate the structure of the disk at 25~\um, but the lack of a high sensitive space based mission in the far-IR makes it impossible to contrast results from \emph{Herschel}.

The most sensitive observations in scattered light close to the star are possibly those presented in \cite{Apai15}. They performed a comprehensive study of \bpic as seen with HST/STIS, including two epochs over 15 years based on which they report no time variations in the brightness or morphology of the disk. 
%They detected the warp in the disk, but were unable to link it to the presence of a secondary disk. 
They recovered the asymmetry between the NE and SW sides in the main plane of the disk, but they didn't find significant evidence of the dust clump, reinforcing the conclusion that the clump cannot be detected in scattered light, only in thermal emission. \cite{Apai15} suggest the asymmetries observed  might be caused by resonances with an unseen planet, or, as we suggest here, a major collision or outgassing event in the disk. Interestingly, when they performed a study of the vertical profile, they found that the peak values of the secondary component left the plane of the main disk in the SW side, towards much higher vertical distances \citep[see Figs. 6, 7 and 9 in][]{Apai15}. This resembles our result for the secondary disk characterization in the MIRI data, where the cat's tail dominates the secondary component at distances larger than $\sim$100~au (5\arcsec). Their Figure 8, showing asymmetries after subtracting an estimate of the main disk, highlights that on the SW side the emission is brighter and wider, but on the NE side faint nebulosity extends outwards from the inclined secondary disk, seen about 5\degr counterclockwise from the main disk plane and visible out to $\sim9$ \arcsec separation. This is spatially coaligned with the inner portion of the more extended ``fork'' seen in the MIRI data, and is consistent with being the inner portion of the eccentric dust disk originating from steady-state collisions at the collision point.
 
The main disk is almost but not precisely edge on, but the inclination appears to vary between disk components at different radii. 
For instance, \cite{Milli2014} reported the presence of a bow in the disk towards the NW, also reported in the optical by \citet{Golimowski06} and \citet{Apai15}, and attributed  to forward-scattering in a not completely edge-on disk. Likewise, in all NIRCam filters the NW side of the inner disk ($\lesssim 3''$ or 60 au) is brighter than the SE side, consistent with the NW side being slightly inclined toward us.  On the other hand, at much wider separations ($\gtrsim 25''$ or 500 au) \citet{Janson21} find the far outer wings of the disk are bowed in the opposite sense, which may require invoking an additional wider disk component or outer warp to explain. 

The understanding we have about the overall structure of the disk invokes now several components, including some already known such as the main disk and a secondary axisymmetric disk responsible for the warp in scattered light; and some new, reported here for the first time, such as the fork, ``cat's tail" and nebulosity, all apparently indicative of a collisional scenario. The asymmetry observed between the NE and SW sides in both MIRI filters, consistent of possibly various clumps ---some of them already reported in the literature--- might also be related to the sporadic dust production model, but a more detailed multi-wavelength analysis is probably required to further investigate its nature, and whether its linked to a collision \citep{Han23} or rather to planetary interactions\citep{Skaf23}.

\subsection{Composition}

\cite{LiGreenberg98} fit a model of the \Bpic disk to photometry and spectra from a broad range of data sets spanning from 2 to 1300 $\mu$m. Because these observations were unresolved, degeneracies between separation, temperature, density, and size distribution were unavoidable in disk-integrated fits. Nonetheless, \cite{LiGreenberg98} concluded that the particles must be very fluffy (potentially 97.5\% porosity) and have an organic refractory mantle surrounding a silicate core to reproduce silicate features as well as the mid-IR flux, broadly consistent with our sporadic dust production model of the cat's tail and secondary disk presented in Sect. \ref{sec:dynamics_model}. \cite{LiGreenberg98} also found that the size distribution must have enhanced small particles compared to comet Halley (including those that contribute to the mid-IR spectrum). While \cite{LiGreenberg98} did not adopt a simple power law size distribution, this roughly translates into $dN/ds\propto s^{-3.1}$ (based on their Figure 2), shallower than the size distribution we invoke for the cat's tail. This difference may be due to the fact that we are fitting an isolated structure, whereas \cite{LiGreenberg98} fit a disk-integrated spectrum.  

Contrary to the results of \cite{LiGreenberg98}, \cite{Golimowski06} modeled resolved scattered light observations of \BPic and found the best fit composition to the main disk to be \emph{non-porous} or moderately porous silicate and graphite grains. They ruled out high porosity on the basis that such grains would be significantly bluer in scattered light than observations. \cite{Ballering16} also investigated the dust composition in \bpic, simultaneously fitting both spatially resolved scattered light \emph{and} thermal emission observations. Similar to \cite{Golimowski06}, their best fit suggests a composition of \emph{non-porous} grains comprised of 60\% astronomical silicates and 40\% organic refractory material. They found no strong evidence for water ice, consistent with our measurement in Sect. \ref{sec:subsec_waterice}. \cite{Ballering16} concluded that the \BPic halo contained sub-blowout sized grains, consistent with the conclusions of \cite{LiGreenberg98} and with recent dust production events like we invoke for the cat's tail.

How can fits to spatially resolved and scattered light data sets conclude no porosity while fits to the MIR SED suggest very high porosity? As Figure \ref{fig:false_color} suggests, the \BPic disk may simply be composed of two distinct populations of grains---one of these may dominate scattered light while the other dominates the mid-IR flux. As modeled in Section \ref{sec:dynamics_model}, the main disk may be comprised predominantly of low porosity grains, while the secondary disk is primarily composed of high porosity organic refractory material. Figure \ref{fig:SEDs_not_normalized} shows the SEDs for porous (blue) and non-porous (orange) organic refractory grains for various grain sizes and circumstellar distances. The 95\% porous grains scatter $\sim300$ times less starlight than their non-porous counterparts near 1 $\mu$m, yet can be $\sim100$ times brighter in the mid-IR. Such grains would be faint in scattered light observations and bright in the mid-IR, potentially explaining the lack of a detection of the cat's tail in scattered light. Indeed, near-IR surface brightnesses predicted for the cat's tail using the SEDs for 95\% porous organic grains are well below the sensitivity limits of the NIRCam observations.  \cite{Golimowski06} concluded that the two disk components must be composed of different grain populations, noted that the secondary disk is tentatively bluer than the main disk, and found that the SW extension of the main disk reddens more quickly than the NE extension, all of which is consistent with our model of an eccentric secondary disk composed of porous organic refractory grains with periastron to the SW. The MIRI bandpasses may provide a unique window into this population of dust.

\begin{figure}
    \centering
    \includegraphics[width=\columnwidth]{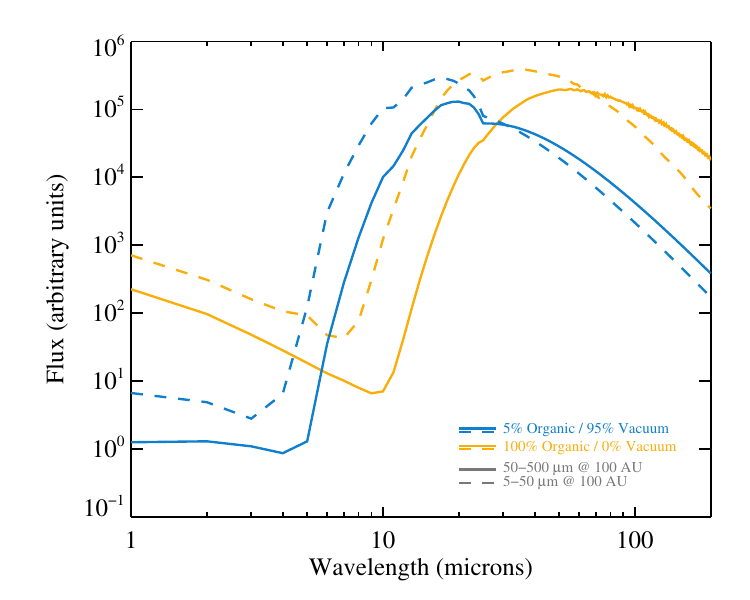}
    \caption{Spectral energy distribution for identical disks with porous (blue) and non-porous (orange) organic refractory material. Porosity reduces the scattered light by a factor of $\sim300$ and increases the MIR flux by a factor of $\sim100$. A secondary disk and cat's tail composed of porous organic refractory material would appear faint in scattered light and bright in the MIR, potentially explaining the relative absence of these features in scattered light as well as the varying best fit compositions appearing in the literature.}
    \label{fig:SEDs_not_normalized}
\end{figure}

\subsection{Origin and evolution}

Possible origins of the secondary disk and its asymmetries have been discussed at length in the literature and have included a recent giant impact, steady-state collision point models, secular forcing by a planet, resonant trapping by a planet, and tidal disruption~\citep[e.g.,][]{Dent2014,Jackson2014,Nesvold2015,Matra2017a,Cataldi18}. Here, we posit that the secondary eccentric disk is indeed generated in a manner consistent with a long-term collision point model, such that much of the CO and C emission may be frequently generated, but the cat's tail and the CO/C \emph{clumps} are the result of an individual recent dust production event superimposed on the secondary disk. We contend that individual, sporadic collisions like what created the cat's tail are the mechanism that populates and maintains the quasi-steady state secondary disk.

\cite{Dent2014} argued against the idea of a recent collision noting the short CO photodissociation timescale within the clump, with literature estimates varying from 50 to 300 years~\citep{Visser2009,Dent2014,Matra2017b,Cataldi18}. Such a short photodissociation timescale implies a very recent collision, which \cite{Dent2014} deemed less likely. Similarly, \cite{Cataldi18} argued against a recent collision based on the short timescale for atomic C production ($\sim5000$ yrs) assuming production primarily via CO photodissociation. However, \cite{Cataldi18} pointed out that the C exhibits the same clumpy nature as the CO, which should have spread out azimuthally within this $\sim5000$ years. 

Our dynamical model of the cat's tail suggests a recent dust production event, with a timescale of just $\sim100$ yrs since production, giving credence to a recent dust production event as the source of the continuum and gas emission clumps. We emphasize that our quasi-steady-state episodic dust production model does not require the entire secondary disk to be produced recently--only the asymmetries/clumps. Our models also suggest abundant organic refractory material in the secondary disk, which may enhance production of C through destruction of carbon-bearing molecules~\citep{Cataldi18}, such that production of C need not arise solely from CO photodissociation. While this material does not necessarily imply a carbon-enriched progenitor, we find it likely that a carbon-enriched progenitor would exhibit more of this material, potentially consistent with the observed over-abundance of C gas~\citep{Roberge2006}. Alternatively, it is possible that the progenitor's bulk composition was not carbon-enriched while the ejected material is, perhaps due to a collision stripping the outer layers from a differentiated body.

What is the probability of such a massive collisional event occurring in the last 150 years? To estimate the probability, we distributed objects among 20k logarithmically-spaced radius bins according to a Dohnanyi size distribution, ranging from 10 microns (roughly consistent with the expected blowout size of the main disk) to 1000 km. 
We normalized our distribution to produce a size-integrated optical depth of $1.7\times10^{-3}$ over a 10\% wide belt centered at 85 au, where the optical depth is taken to be equal to $L_{\rm IR} / L{\star}$ \citep{Kral2017}. 
For each bin of radius $s$, we then calculated the collision time $t_{\rm coll}\!\left(s\right)= t_{\rm orbit}/\left(4 \pi \tau'\!\left(s'\right)\right)$ \citep{Wyatt1999}, where $t_{\rm orbit}$ is the orbital period at 85 au, $\tau'\!\left(s'\right)$ is the optical depth of bodies larger than than $s'=s/2$, and $s'$ is the smallest object capable of fragmenting an object of radius $s$. 
The probability of collision for any single body within the past 150 years can be approximated using the Binomial theorem as $p\!\left(s\right) = 150\;{\rm yr}/t_{\rm coll}\!\left(s\right)$. 
The probability of at least one collision occurring in the past 150 years among all bodies larger than the minimum progenitor is approximately given by

\begin{equation}
    \sum_{s=s_{\rm min}}^{\infty} 1 - \left(1-p\!\left(s\right)\right)^{N\!\left(s\right)},
\end{equation}
where $N\!\left(s\right)$ is the number of objects in our bin of radius $s$.

In Section \ref{sec:sporadic_model} we estimated the mass of the progenitor to be at least $10^{19}$ kg and potentially as large as $10^{21}$ kg. In the solar system, these two masses correspond roughly to radii of $\sim100$ km and $\sim500$ km, respectively. Adopting these radii for $s_{\rm min}$, we find a 65\% probability of at least one collision of a body larger than 100 km in radius occurring in the past 150 years, and a 0.5\% probability of at least one collision of a body larger than 500 km in radius. 

We stress that estimates of the probability of collisional events are wildly uncertain. The estimate above must rely on extrapolations of a power law size distribution over many orders of magnitude in size. Small changes in this power law can greatly affect the estimated probability. Small changes in the assumed blowout size that helps benchmark the optical depth can also greatly affect the assumed probability. Precise estimates would require knowledge of the absolute number and orbits of large planetesimals, for which we have no knowledge. As a result, these estimated probabilities come with unquantifiable, but large, uncertainties. We therefore conclude that the calculated probabilities are consistent with our model, in that they do not clearly rule out such a scenario.

\Bpic's large number of comets---in fact, the largest of any exoplanetary system studied thus far---might also tie in to our proposed dynamical scenario. The exocomets were first detected as variable absorption features in the spectra of \Bpic due to refractory materials \citep{Ferlet87,Lagrange-Henri89}. The high frequency of exocomet events led to several studies trying to find an origin \citep{Beust90,Beust91}, which quickly linked the presence of a perturbing body in the disk \citep{BeustMorbidelli00,ThebaultBeust01}. The planet \bpic b, detected years later \citep{Lagrange09}, seemed a natural match to the planetary perturbation hypothesis. Through a study of hundreds of exocometary transits, \cite{Kiefer14} found that two families of exocomets can be distinguished in orbit around \bpic. They attribute the origin of one of the families, named \emph{S} due to shallow transits, to the resonances caused with planet b. However, family \emph{D} comets, with deeper transits and larger periastrons, have a narrow distribution of periastron longitudes, and therefore very similar orbits. This might indicate that they originate from the disruption of one or several individual larger exocomets or parent bodies. A recent major dust production event or events could plausibly send debris inward toward the star as well as outwards. Whether family \emph{D} may be linked at all to a similar scenario such as the one described in Sect. \ref{sec:dynamics_model} will require a more in depth analysis. 

Our sporadic dust production model suggests a cat's tail comprised of $\sim100$ K dust with $\beta\sim10$. Such a model leads to several testable predictions. First, the spectra of the cat's tail should reveal an SED similar to the blue curves (for porous organic grains) shown in Figure \ref{fig:SEDs}, and the variation of this spectrum over the length of the cat's tail should suggest smaller grains near the tip. Second, the high $\beta$ value suggests that dust in the mid-to-outer regions of the cat's tail, in spite of being located at circumstellar distances approaching $\sim1000$ au, should be fast moving. We predict a $\sim200$ mas shift in the location of the cat's tail over three years. Finally, our model predicts that the secondary disk should not be perfectly edge-on, such that larger grains should track an ellipse with a semi-minor axis $\sim0.2$\arcsec, which should be resolvable with ALMA.

\section{Conclusions} \label{sec:conclusions}

The JWST high contrast images of the \bpic disk presented here have proven to be a crucial addition to the long effort to characterize and understand this complex system. The unprecedented sensitivity along with high spatial resolution in near and mid-IR wavelengths allow for a better understanding of the dust distribution along the disk, clearly revealing the two disk components already inferred in previous works and detecting several new components. 
In this work, we have reported:
\begin{itemize}
    \item The presence of two distinct disk components, with the secondary component having a relative position angle to the main disk of 5.6$\pm$0.4~$^\circ$. Moreover, the main and secondary disk exhibit different colors in MIRI's bandpasses, suggesting the material in the secondary disk is hotter, and therefore presumably a different composition than the main disk. We also identified a ``fork" that only appears on the NE side of the disk extending beyond 15\arcsec, appearing in MIRI observations and attributed to this secondary disk. This inner portion of this secondary component has been observed before in scattered light at similar stellocentric distances, appearing as an apparent warp in the disk. Composite color images of the MIRI observations confirm the secondary disk is eccentric, with periastron to the SW, and is comprised of hotter grains.  NIRCam observations are consistent with previous scattered light studies of the disk.
    \item A curved asymmetric feature to the SW side of the disk, which we dub the ``cat's tail." This feature exhibits the same F1550C/F2300C color as the secondary disk and seems to have its origin at the location of the CO clump and mid-IR asymmetry. The cat's tail is only detected in mid-IR, i.e. thermal emission. This is consistent with porous organic refractory grains, which can appear bright in the MIR and faint in scattered light. Dedicated spectroscopic and/or imaging studies are needed at additional wavelengths in order to further investigate the dust composition and size distribution.
    \item Tendrilous nebulosity to the SW and W of the disk. These arc-like features also exhibit the same F1550C/F2300C color as the secondary disk.
    %\item No exceptional color variation along the main disk. 
    
\end{itemize}

We presented a sporadic dust production model dynamically linking the eccentric secondary disk to the cat's tail and tendrilous nebulosity observed to the SW. We posited that the eccentric secondary disk is due to ancient or ongoing dust production, similar to the steady state collision point model of \cite{Jackson2014}, but the cat's tail and nebulosity, as well as the previously observed mm wavelength excess and CO clump~\citep{Dent2014}, are the outcome of individual recent dust production events occurring at locations that deviate from the collision point. To reproduce the curvature of the cat's tail, we require the production of abundant sub-blowout grains produced within the last $\sim100-200$ years, which is on par with the CO photodissociation timescale.

We find that porous organic refractory material could simultaneously explain the shape of the cat's tail, the color of the secondary disk in the MIR, the flux along the extent of the cat's tail in the MIR, the apparent absence of these features in scattered light, and potentially the previously detected clump of atomic C~\citep{Cataldi18}. This material has been previously invoked to fit the spectral energy distribution of the \bpic disk and suggests a cometary origin~\citep{LiGreenberg98}. JWST's MIRI instrument may provide the unique capability to characterize this population of dust. In-situ investigations of small bodies in the Solar System, such as OSIRIS-REx, will also provide new information about composition of asteroids and comets, and serve as a baseline for research on debris disks composition. 

If the material in the disk has been stirred, it could also explain the large influx of small sized bodies observed in the system as exocomets. While exocomets have been previously studied as originated due to resonances with the planets, new dynamical studies are needed to better understand their origin.

Future observations from both ground and space should aim to expand the wavelength and time regime in which the cat's tail is detected, in order to characterize the material and better constrain the origin of the collisional event(s). The proposed model makes testable predictions for the dust SED along the cat's tail region and for proper motion over time of the features seen with MIRI. 

\begin{acknowledgments}
The authors are grateful for the many efforts of the JWST observatory operations teams, in particular for the flight operations team members who responded to the spacecraft anomaly and payload safing event which occurred during the first attempt at these NIRCam observations, diagnosed the problem, and returned JWST to science operations in the midst of the holiday season and shortly before the one year anniversary of its launch. The authors thank Alycia Weinberger, Aki Roberge, Stefanie Milam, and Nathan Roth for insightful conversations while attempting to model this disk. 
This paper reports work carried out in the context of the JWST Telescope Scientist Team (\url{https://www.stsci.edu/~marel/jwsttelsciteam.html}) (PI: M. Mountain). Funding is provided to the team by NASA through grant 80NSSC20K0586.
Based on observations with the NASA/ESA/CSA JWST, obtained at the Space Telescope Science Institute, which is operated by AURA, Inc., under NASA contract NAS 5-03127. 
The data presented in this article were obtained from the Mikulski Archive for Space Telescopes (MAST) at the Space Telescope Science Institute. The specific observations analyzed can be accessed via \dataset[DOI]{https://www.doi.org/10.17909/8yq1-qv46}.
These observations are associated with JWST program 1411 (PI Stark).

I.R. is supported by grant FJC2021-047860-I and PID2021-127289NB-I00 financed by MCIN/AEI /10.13039/501100011033 and the European Union NextGenerationEU/PRTR.
\end{acknowledgments}

\vspace{5mm}
\facilities{JWST}

\software{python, astropy, spaceKLIP %astropy \citep{2013A&A...558A..33A,2018AJ....156..123A},  
          }

\bibliography{main}{}
\bibliographystyle{aasjournal}

%\appendix
\newpage
%\section{Appendix}
\appendix

\section{Deconvolution Validation}\label{app:deconv_validation} 
To provide a preliminary validation of the utilized deconvolution technique, we applied it to a convolved synthetic disk model having a known initial (unconvolved) flux distribution that could be used as ``ground truth".  For this purpose, we generated an oversampled (by a factor of two) model image for a \bpic-like disk.  For each roll angle of the data, we then rotated and convolved this model for the NIRCam F444W filter using the same procedure as for MCRDI (Section \ref{sec:mcrdi}), except a) using a finer grid of 81 PSFs (10 radial positions, 8 azimuthal positions, and the origin), and b) using an OPD map from a month prior to the observations. Respectively, these distinctions help to insure that the convolved model includes the majority of the spatial PSF variations and that the PSF used for deconvolution is not a perfect match for the PSFs used to convolve the data. After convolution, we downsampled each roll's image to match the detector sampling. Finally, using the same PSF as for the real data, we performed deconvolution on this synthetic dataset as described in Section \ref{sec:deconv}. As a comparison, we also carried out deconvolution using the standard Richardson-Lucy algorithm without consideration for the coronagraph. The results are shown in Figure \ref{fig:deconv_validation}.

%\mdp{Kellen can you please add here one or two more sentences to explain and interpret the results. State to readers how/why the results in Fig 7 validate the deconvolution method.}

These results show that both standard R-L deconvolution and the coronagraphy variant are able to recover the ground truth surface brightness beyond $\sim 2\arcsec$. At smaller separations, the R-L variant significantly outperforms the standard R-L method. This experiment demonstrates that deconvolution assuming a fixed PSF can be suitable for application to such data despite the truly spatially varying PSF. The remaining inaccuracy of the adopted R-L variant at small separations is largest in the filter presented; e.g., for F210M, measurements match the ground truth to within $\sim5\%$.

\begin{figure*}
    \centering
    \includegraphics[width=0.75\textwidth]{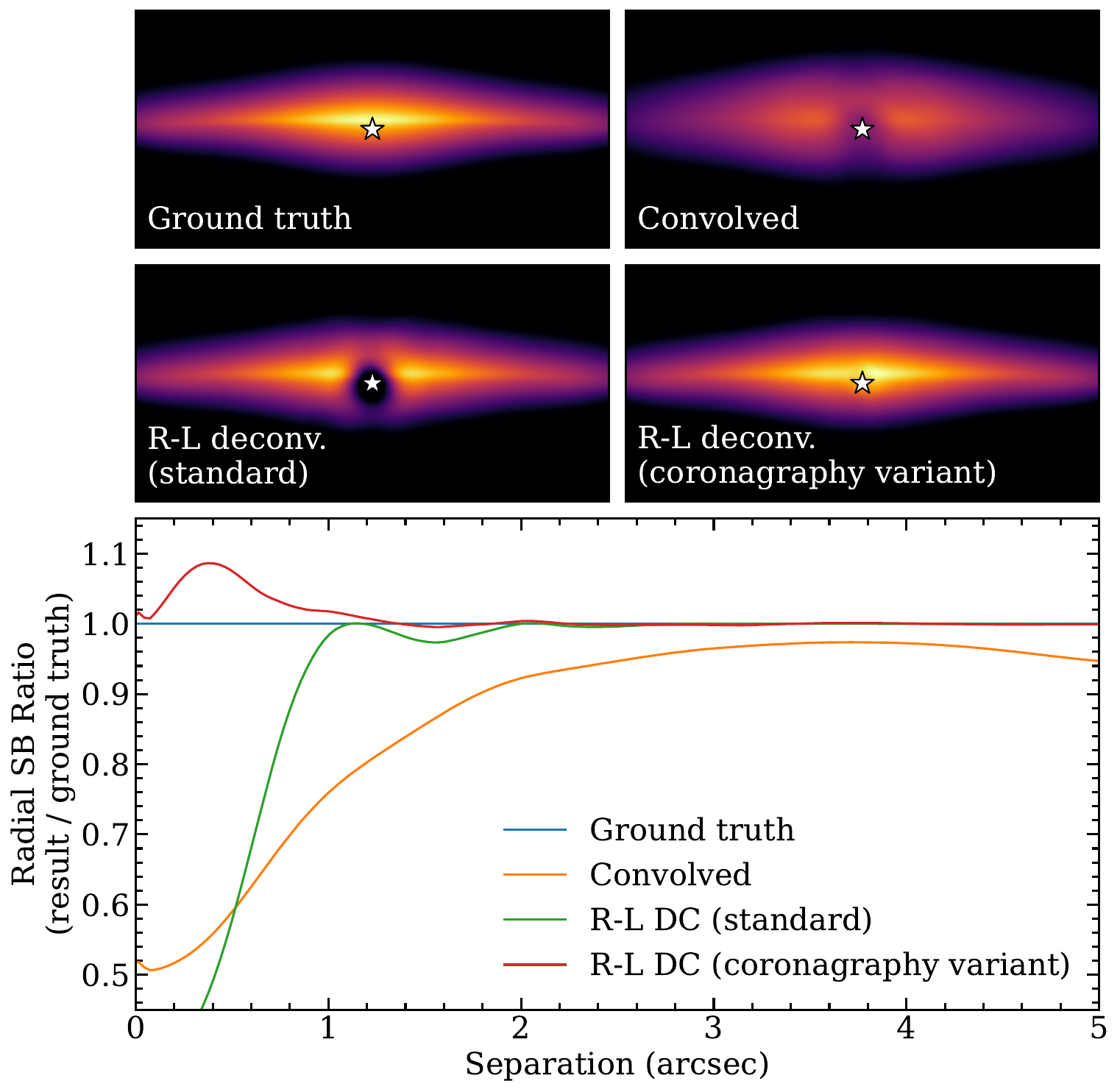}
    \caption{Numerical experiment validating the deconvolution method used. Top row: the ``ground truth'' synthetic disk image (left) and the corresponding PSF-convolved image as described in Appendix \ref{app:deconv_validation}. Middle row: the results for standard Richardson–Lucy (R–L) deconvolution (left), and the R–L variant that considers coronagraphic transmission (right). All images are shown with a logarithmic stretch and a $12\arcsec \times 6\arcsec$ field of view. Bottom panel: the ratio of radial surface brightness measurements for each of the images of the upper panel with those of the ground truth. Surface brightness measurements are made using one pixel wide rectangular apertures that extend to the edge of the field of view. Measurements shown are for the right side of the disk, but values for the left side are comparable. The R-L variant for coronagraphy does far better at correctly reconstructing the disk surface brightness within the inner arcsecond.}
    \label{fig:deconv_validation}
\end{figure*}

\section{Secondary disk fitting}

\label{app:secondary_disk}

Figures showing the result of the fitting of the secondary disk for both NIRCam and MIRI are shown in this section, following the methods described in section \ref{sec:component_fits}. 

\begin{figure}
    \centering
    \includegraphics[width=0.66\columnwidth]{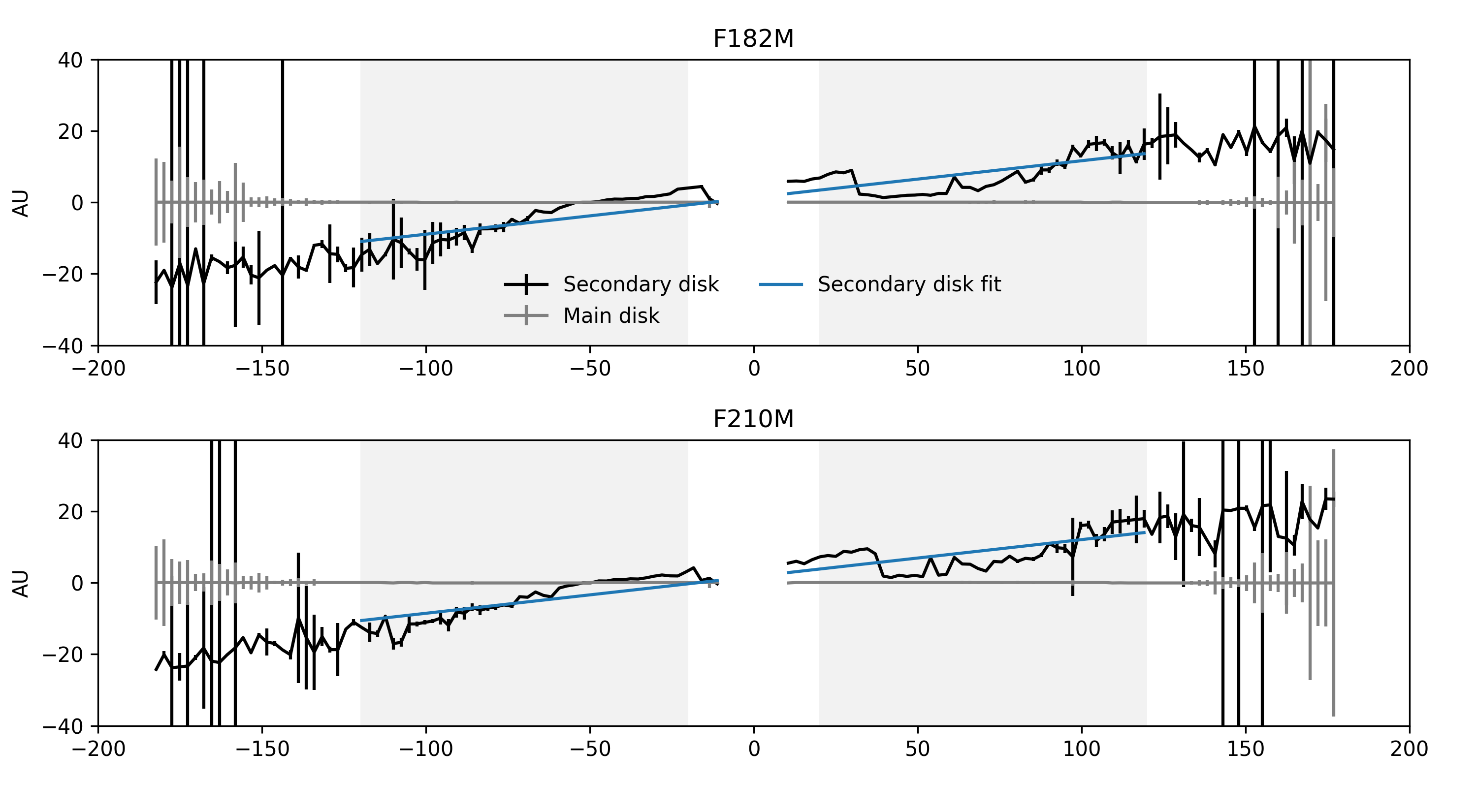}
    \includegraphics[width=0.66\columnwidth]{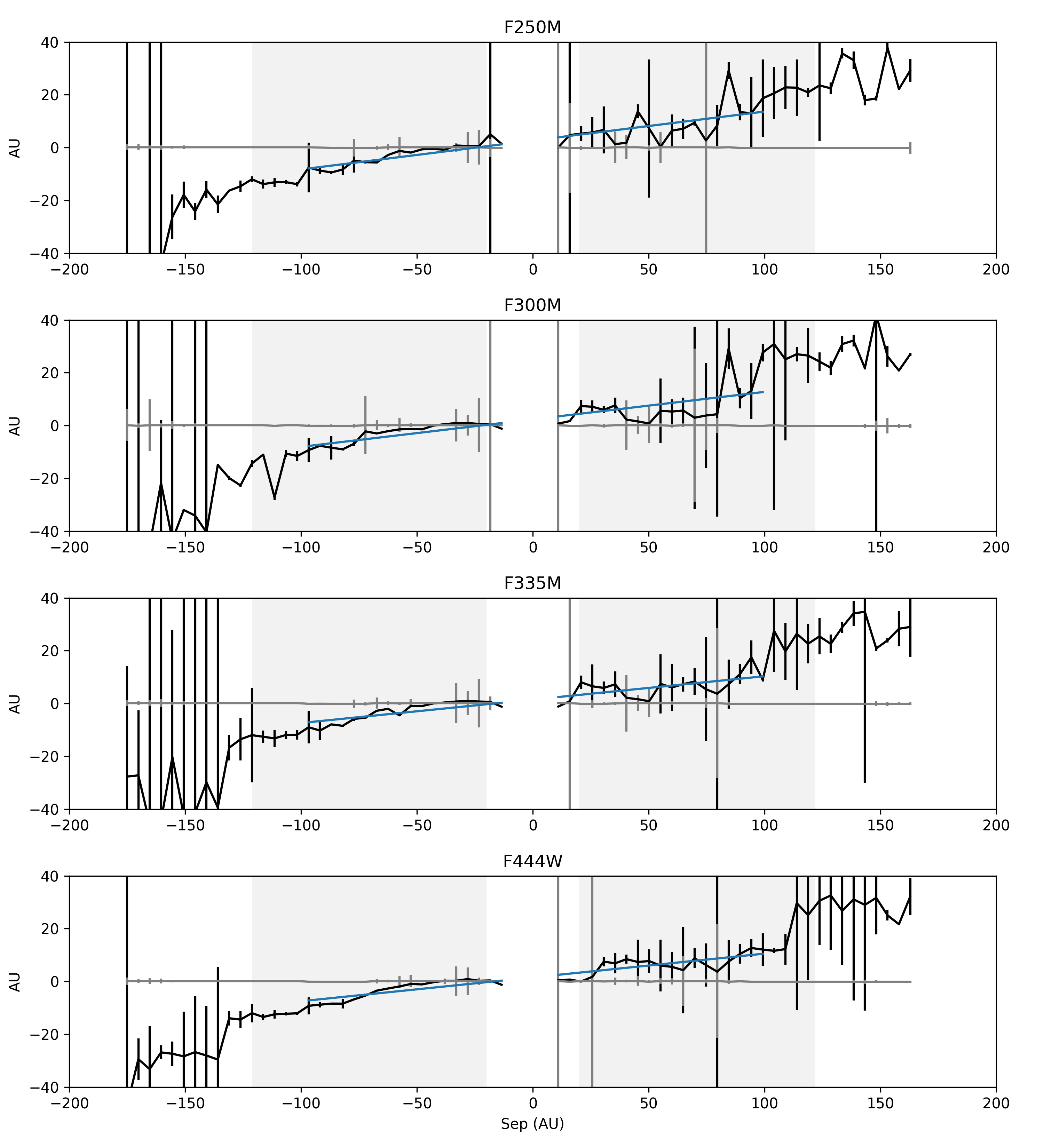}
    
    \caption{Fitting of the location of the secondary disk for the NIRCam short and long filters, using a multi component fit following the approach of \citet{Golimowski06}, their figure 9. Shaded area shows the data used in the fit, to avoid those with large errors in the determination of the main disk (i.e., low SNR of the detection). }
    \label{fig:SD_nircam}
\end{figure}

\begin{figure}
    \centering
    \includegraphics[width=0.66\columnwidth]{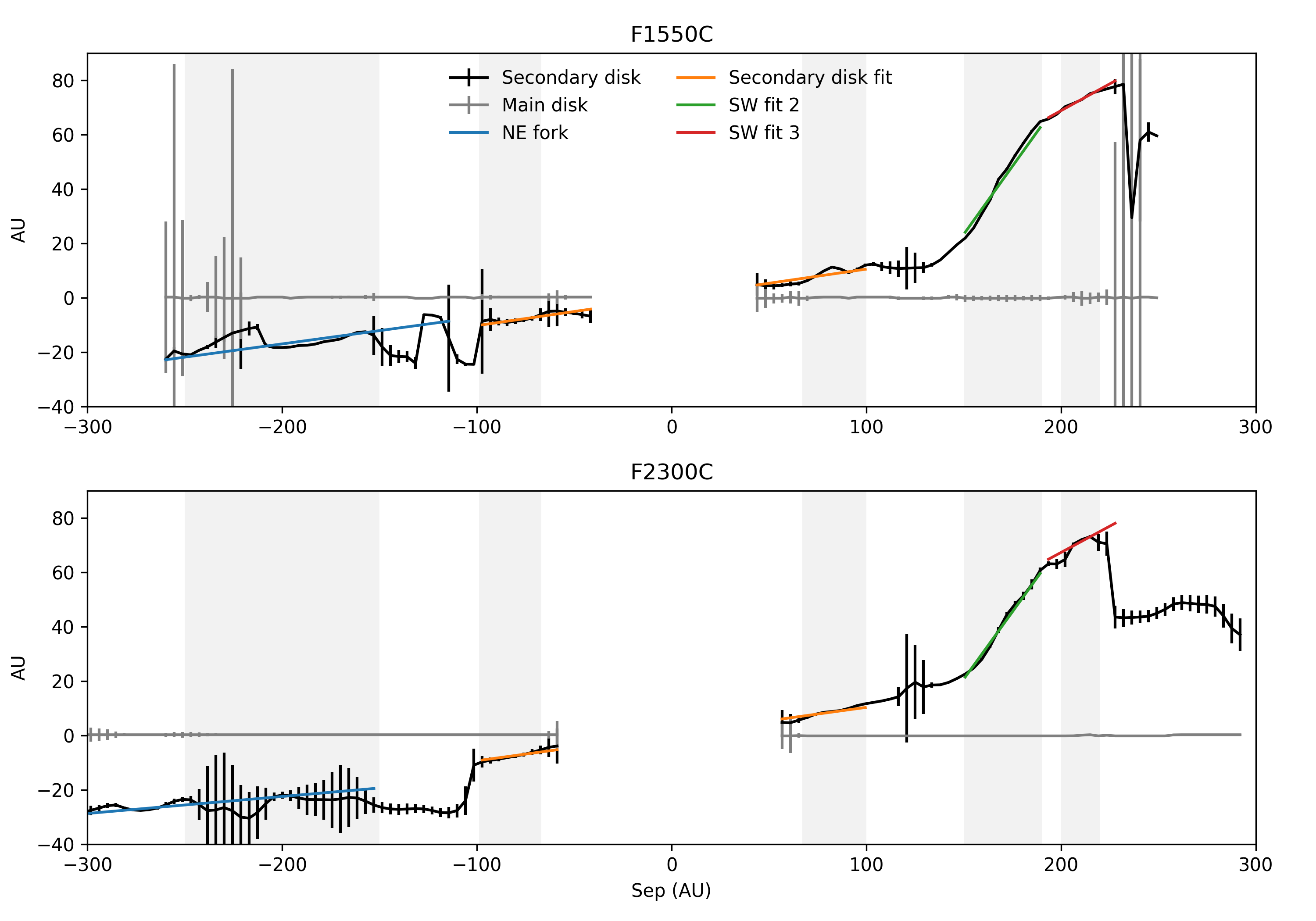}
    
    \caption{Same as Fig. \ref{fig:SD_nircam} but for MIRI filters. Instead of a single fit for the secondary disk, in this case we fitted the NE and SW separately, and further divided the SW side in 2 sections: (1) base of the cat's tail; (2) tip of the cat's tail. FOV are slightly different for F1550C (24\arcsec) vs MIRI (30\arcsec), leading to different spatial coverages. }
    \label{fig:SD_miri}
\end{figure}

\end{document}